\documentclass[journal=jacsat,manuscript=article]{achemso}
\usepackage{achemso}
\usepackage[version=3]{mhchem}
\usepackage[english]{babel}
\usepackage{filecontents}
\usepackage{graphicx}
\graphicspath{{./Figures/}}
\usepackage{courier}
\usepackage{amssymb,amsmath,amsthm,mathrsfs,comment,afterpage,accents}
\usepackage{mathtools}
\usepackage{braket}
\usepackage{mathrsfs}
\usepackage{courier}
\usepackage{multirow}
\usepackage{tabu}
\usepackage{color}
\usepackage{caption}
\usepackage{subcaption}
\usepackage{xcolor}
\usepackage[numbers]{natbib}
\usepackage{xr} 
\usepackage{soul}
\usepackage[normalem]{ulem} 

\makeatletter
\def\@dotsep{4.5}
\makeatother
\usepackage{dcolumn}
\usepackage{bm}
\renewcommand\vec\mathbf

\setkeys{acs}{maxauthors = 0}

\newcommand{\insertnew}[1]{{\textcolor{black} {#1}}} 

\newcommand{\replacewith}[2]{\textcolor{black}{#2}} 

\newcommand{\remove}[1]{}  

\newcommand{\Insertnew}[1]{{\textcolor{black} {#1}}} 

\newcommand{\Replacewith}[2]{\textcolor{black}{#2}} 

\newcommand{\Remove}[1]{}  

\newcommand{\eng}[1]{{\textcolor{black} {#1}}} 

\title{Making Many-Body Interactions Nearly Pairwise Additive: The Polarized Many-Body Expansion Approach}
\author{Srimukh Prasad Veccham}
\author{Joonho Lee}
\email{linusjoonho@gmail.com}
\author{Martin Head-Gordon}
\email{mhg@cchem.berkeley.edu}
\affiliation{
Department of Chemistry, University of California, Berkeley, California 94720, USA
Chemical Sciences Division, Lawrence Berkeley National Laboratory, Berkeley, California 94720, USA
}

\begin{document}
\maketitle
\newpage
\begin{abstract}
\insertnew{The} \replacewith{Many-body expansion}{Many-Body Expansion} (MBE) is a useful tool to simulate condensed phase chemical systems, often avoiding the steep computational cost of usual electronic structure methods.  
However, it often requires\remove{a} higher\remove{order} than 2-body terms to achieve quantitative accuracy.  
In this work, we propose the \replacewith{polarized}{Polarized} MBE (PolBE) method where each MBE energy contribution is treated as an embedding problem.
In each energy term,  a smaller fragment is embedded into a larger,\insertnew{ } polarized environment and only \replacewith{the}{a} small region\remove{in the system} is treated at the high-level of theory using embedded mean-field theory. 
The role of polarized environment was found to be crucial in providing quantitative accuracy at the 2-body level.  
PolBE accurately predicts non-covalent interaction energies for a number of systems, including \ce{CO2}, water, and hydrated ion clusters, with a variety of interaction mechanisms, from weak dispersion to strong electrostatics considered in this work.  
We further demonstrate that the PolBE interaction energy is \replacewith{mainly}{ predominantly}  pairwise unlike the usual vacuum MBE which requires\remove{a} higher-order\remove{interaction} term\insertnew{s} to achieve\remove{a} similar accuracy.  
We numerically show that PolBE often performs  better  than  other  widely  used  embedded  MBE  methods  such  as  the  electrostatically  embedded  MBE.  
Owing  to  the  lack  of  expensive  diagonalization  of  Fock matrices and its embarrassingly parallel nature,  PolBE is a promising way to access condensed  phase  systems  with  hybrid  density  functional\insertnew{s}\remove{theory}  that  are  difficult  to treat with currently available methods.
\end{abstract}
\newpage

\section*{Introduction}
Simulating condensed phase systems is considered the holy grail of quantum chemistry. 
However historically, the majority of quantum chemistry research has focused on molecular properties rather than bulk properties of liquids. 
The major hurdle towards calculating and predicting properties of bulk materials is the steep computational scaling in modeling them.
For decades, a major research goal of the quantum chemistry community has been to formulate more accurate and inexpensive computational methodologies to predict properties of bulk materials.

For systems comprising of non-covalently bonded fragments, the many-body expansion method has been used as way to circumvent the steep hyper-linear scaling of quantum chemistry methods.
The \replacewith{many-body expansion}{Many-Body Expansion} (MBE) is a technique for calculating the properties of a large system by (1) partitioning the system into $N$ fragments, (2) calculating fragment properties, and (3) reassembling them using the many-body expansion equation.
This approach reduces a large supersystem calculation into a large number of much smaller fragment calculations, and each of them is completely independent of others. Thereby, MBE calculations are ``embarrassingly parallel.''
While the MBE is in principle exact, it provides computational advantages only when it is truncated up to a certain order.
Truncated many-body expansions are not exact\eng{,} and  its accuracy can heavily depend on the underlying chemical interactions in the system.
Some applications of the many-body expansion method are determination of lattice energies \cite{Yang2014} and studying polymorphism in molecular crystals. \cite{Beran2016}
Many-body based cluster corrections to calculations with periodic boundary conditions have also served as powerful tool\insertnew{s} to study crystal structures using highly accurate wavefunction techniques. \cite{Boese2017}
\insertnew{MBEs have also been used to understand and incorporate the effects of mutual polarization in classical molecular dynamics force fields.\cite{Demerdash2014}}

The convergence of MBE can be accelerated using embedding, in which each $k$-body term is calculated in the presence of the remaining $(N-k)$ fragments (hereafter referred to as the `environment').
An active research question is then how one models this environment and its interaction with the $k$-body term at a given electronic structure level.
Quantum embedding can be performed in the context of Green's functions, \cite{Zgid2011,Lan2017, Ishida2001, Chibani2016} densities or potentials,\cite{Elliott2010,Huang2016,Huang2011,Fornace2015,Manby2012, Wesolowski1993, Wesolowski2015} or density matrices. \cite{Knizia2012,Wouters2016}
In many chemical phenomena, the interesting chemical activity like binding or chemical reaction occurs only in a small part of the chemical system.
\eng{Embedding techniques, taking advantage of this ``nearsightedness''}\cite{Prodan2005} in order to \eng{divide the \replacewith{molecule of study}{system} into two parts: A - the \replacewith{system}{region} of interest where an actual chemical phenomenon occurs and B - the environment which surrounds the system but does not directly take part in the actual chemical process.}
Even though the environment does not directly participate in the chemical phenomenon, it sufficiently modulates the chemical process occurring in the system that it cannot be neglected.
Embedding can also be done by modelling the environment using molecular mechanics which constitute widely used QM:MM methods like ONIOM.\cite{Vreven2006}
ONIOM and its extensions \cite{Guo2010} have been extensively used to study a wide range of systems and phenomena from heterogeneous catalysis to biological macromolecules. \cite{Chung2015}
\Insertnew{However, ONIOM does not contain any explicit system-environment coupling terms in the hamiltonian, and cannot accurately capture the effects of environment.}
Embedded MBEs (eMBE) make use of quantum embedding techniques to describe the $k$-body term using an accurate and often more expensive computational method while using a much cheaper method to represent the environment.
An eMBE truncated at the 2-body level aims to reduce the relative importance of the higher order terms (3-body and higher) by \eng{folding in} these many-body effects into the 1-body and the 2-body terms.
Different models of the environment, which can capture distinct physical effects, are capable of encoding these many-body effects with varying degrees of accuracy and computational expense.

A simple way to go beyond vacuum MBE, where we do not include any environment, is to model the environment\Remove{is} using atom-centered point charges, termed the Electrostatically-Embedded Many-Body Expansion (EE-MBE). \cite{Dahlke2007a, Dahlke2008a, Dahlke2007}
These atom-centered point charges incorporate some effect of electrostatic interaction while the finite-\replacewith{size}{extent}  of electron density, exchange interaction, and polarization of the environment are neglected.
EE-MBE is fairly simple to implement, however its accuracy is very dependent on the actual charges used for embedding. \cite{Richard2014}

\Insertnew{One way to go beyond a simple charge embedding model is to use the Embedded Many-Body Expansion of Manby and co-workers which uses atom-centered gaussians to represent electrostatic interactions of the environment and a simple empirical model for Pauli repulsion. \cite{Bygrave2012}}
A slightly more sophisticated embedding method is the Variational Many-Body Expansion (VMBE), \cite{Gao2012} which builds upon \Replacewith{a cruder}{the 1-body} X-Pol wavefunction, \cite{Song2009, Gao1996, Gao1998,Xie2007, Liu2019VariationalDynamics, Lao2012AccurateDispersion} introduced by Jiali Gao and co-workers. 
In the X-Pol wavefunction, the supersystem wavefunction is written as a Hartree-product of the monomer wavefunctions. 
Inter-fragment coulombic interactions are simplified by using point charges in order to avoid computing two-electron integrals for the whole system. 
The X-Pol wavefunction can either be optimized variationally or by using a double self-consistent field procedure. \cite{Song2009}
Dispersion, exchange repulsion, and charge transfer effects are approximated using empirical Lennard-Jones terms.
\Insertnew{A variation of the X-Pol wavefunction, X-Pol-X, can be used to incorporate the effect of exchange repulsion non-empirically. \cite{Cembran2010}}
While the X-Pol wavefunction treats these interactions empirically, VMBE makes an attempt to treat them quantum mechanically at the 2-body level by redefining the fragment in the X-Pol wavefunction to contain two monomers.
It is important to note that the environment density is updated for each of the dimer calculations in order to include complete mutual polarization between the dimer and its environment.
The Fragment Molecular Orbital (FMO) method is another idea based on the MBE, which uses densities obtained from monomer wavefunctions to represent the fragments in the environment. \cite{Kitaura1999,Fedorov2007, Gordon2012, Fedorov2006}
It has been identified that the right choice of a zeroth-order wavefunction, a wavefunction that can be used to form an inexpensive model for the environment, can lead to significantly accelerated convergence of the many-body expansion. \cite{Manby2012, Gao2012}
A good zeroth-order wavefunction is one that is able to incorporate the most important interactions present in the system in a computationally efficient manner.

Our choice for the zeroth order wavefunction is the Self-Consistent Field for Molecular Interaction (SCFMI) wavefunction, which was initially introduced as an efficient representation for weakly interacting systems. \cite{Stoll1980, Gianinetti1996, Khaliullin2006, Nagata2001BasisFormalism,Cullen1991}
SCFMI utilizes localized molecular orbitals and enforces fragment sparsity in the molecular orbitals.
In SCFMI, molecular orbitals are expanded in local subsets of atomic orbitals and are called Absolutely Localized Molecular Orbitals (ALMOs). \cite{Stoll1980}
Like other localized molecular orbitals, ALMOs can be used in chemical interpretation \cite{Khaliullin2007, Horn2013, Thirman2015AnOrbitals, Mo2000} and formulation of reduced scaling methods using local correlation techniques.\Remove{\cite{Bygrave2012}}

Partitioning of AOs into subsets naturally arises for systems consisting of non-covalently bound entities like molecular clusters.
ALMOs are orthogonal on fragment but non-orthogonal between fragments.
ALMOs have been routinely used to perform interaction energy calculations which are basis set superposition error free,\cite{Gianinetti1996} employed to carry out computationally cheaper SCF alternative for weakly interacting systems, \cite{Khaliullin2006} and used to decompose interaction energy in energy decomposition analysis (i.e. ALMO-EDA) schemes. \cite{Khaliullin2007,Horn2013,Thirman2015AnOrbitals,Mo2000}
Unlike SCF, SCFMI avoids diagonalization of the full system Fock matrix which reduces the prefactor for linear algebra in comparision to an SCF calculation.

Moving forward using eMBE with SCFMI, another computational bottleneck will appear for Density Functional Theory (DFT) calculations.
The most accurate hybrid Exchange Correlation (XC) functionals involve some amount of exact exchange, which in turn becomes a bottleneck for large-scale calculations.\cite{Mardirossian2017}
One way to circumvent this is to further approximate the environment beyond SCFMI.
We have noticed that it is sufficient to capture only the important interactions between each of the computed terms and its surrounding in order to ensure expedited convergence of the many-body expansion.
In this paper, we represent the 1-body and 2-body terms using hybrid density functionals with appropriate dispersion corrections because of their ability to accurately represent non-covalent interactions.
The environment is represented using a semi-local functional due to its cheaper computational cost.
The Embedded Mean-Field Theory (EMFT) developed by Manby and Miller is well-suited for embedding a hybrid density functional in a semi-local density functional. \cite{Fornace2015}
EMFT is not only simple to implement, but it has also been shown to reduce the cost of exact exchange computation while maintaining accuracy. \cite{Fornace2015}
EMFT treats the system and its environment as open quantum systems incorporating quantum entanglement at a mean-field level.
The combination of eMBE, SCFMI, and EMFT will be referred to as PolBE. 
This paper is organized as follows:
the theoretical background behind the MBE \eng{will} be introduced.
ALMOs are defined and the procedure to obtain them self-consistently, namely SCFMI, will be described.
The EMFT method and models for computing exact exchange will be presented.
The performance of PolBE for several chemical systems will be assessed in comparison with other existing eMBE methods.
Finally, we will discuss the computational cost of a preliminary implementation of PolBE.

\newpage
\section{Theory}

\subsection{Embedded Many-body Expansion (eMBE)}
The Many-Body Expansion (MBE) for a system of $F$ fragments is written as a sum of 1-body ($V_1$) , 2-body ($V_2$), ... , $F$-body ($V_F$) terms as shown in Eq.~\eqref{eq::MBE}.

\begin{align} \label{eq::MBE}
    E_\text{total} &= \sum_{i=1}^{F} V_i
\end{align}
where
\begin{align}
    V_1 &= \sum_{i=1}^{F} E_i(B),     \label{eq::1body} \\
    V_2 &= \sum_{i<j}^{F} (E_{ij}(B) - E_i(B) - E_j(B)),      \label{eq::2body} \\
    \begin{split}
            V_3 &= \sum_{i<j<k}^{F} [(E_{ijk}(B) - E_i(B) - E_j(B) - E_k(B)) \\ 
            &- (E_{ij}(B) - E_i(B) - E_j(B)) - (E_{ik}(B) - E_i(B) - E_k(B)) - (E_{jk}(B) - E_j(B) - E_k(B))]  \label{eq::3body}
    \end{split}
\end{align}
and so on, $E_i(B)$ is the energy of the $i$-th monomer computed in the presence of the environment denoted as `B', $E_{ij}(B)$ is the energy of dimers $(ij)$ computed in the presence of the environment, and $E_{ijk}(B)$ is the energy of the trimers $(ijk)$ computed in the presence of the environment, and so on.

Different eMBE methods differ only in their treatment of the environment.
For example, in the Electrostatically-Embedded Many-Body Expansion, the 2-body term is $E_{ij}(B)$ is calculated in the presence of atom-centered point charges for all monomers $k \neq i,j$. 
This embedding accelerates the convergence of the MBE, in comparison to vacuum MBE in which there is no environment.
PolBE represents the environment using the polarized SCFMI density.
In this paper, unless otherwise stated, all the MBEs and the eMBEs have been truncated at the second order, applying the pairwise approximation.


\subsection{SCFMI model for the environment}
Absolutely Localized Molecular Orbitals (ALMOs) on a fragment X ($| \psi _{\text{X}i} \rangle$) are linear combinations of atomic orbitals belonging to the same fragment ($| \phi_{\text{X}\mu} \rangle$) as shown in Eq.~\eqref{eq::ALMO}. 
\begin{align}
    | \psi _{\text{X}i} \rangle = \sum _{\mu} | \phi_{\text{X}\mu} \rangle C_{\bullet \text{Xi}}^{\text{X}\mu \bullet} \label{eq::ALMO}
\end{align}
where \textbf{C} is the MO coefficient matrix, in which all the MO coefficients $C_{\bullet \text{Xi}}^{\text{Y}\mu \bullet} $ when $X \neq Y$ are constrained to be zero (i.e. fragment sparsity), giving the \textbf{C} matrix a block diagonal structure.

Solving for the MOs with the ALMO constraint gives the SCFMI equations for each fragment X as shown in Eq.\eqref{eq::Projected_Fock}.
\begin{align}
    [\textbf{F}]_{\text{XX}} [\textbf{C}]_{\text{XX}} &= [\textbf{S}]_{\text{XX}}[\textbf{C}]_{\text{XX}}[\epsilon]_{\text{XX}} \label{eq::Projected_Fock} \\
    [\textbf{F}]_{\text{XX}} &= (\textbf{I}-\textbf{SP}+\textbf{SP}^{\text{X}})\textbf{F}(\textbf{I}-\textbf{PS}+\textbf{P}^{\text{X}}\textbf{S}) \label{eq:Projected_Fock_formation} \\
    \big(\textbf{P}^{\text{X}}\big)^{Z\nu, X\mu} &=  (\textbf{T})^{Z\nu \bullet}_{\bullet Zl} ( \boldsymbol\sigma^{-1})^{Zl, Xj} (\textbf{T}^T)^{\bullet X\mu} _{Xj\bullet} \label{eq:Stoll_projector} 
\end{align}
where $[\textbf{F}]_{\text{XX}}$ is the projected Fock matrix constructed as shown in \eqref{eq:Projected_Fock_formation}, $[\textbf{C}]_{\text{XX}}$ is the ALMO coefficient matrix, $[\textbf{S}]_{\text{XX}}$ is the fragment AO overlap matrix, and $[\epsilon]_{\text{XX}}$ contains the \insertnew{fragment} orbital energies.
The projected Fock matrix ($[\textbf{F}]_{\text{XX}}$)\remove{is}  for each fragment is constructed using the density matrix ($\textbf{P}$), full system AO overlap ($\textbf{S}$) and, the Stoll projector $\textbf{P}^{\text{X}}$ as shown in Eq.~\eqref{eq::Projected_Fock}.
$\textbf{T}$ is the occupied orbital coefficient matrix and $\boldsymbol\sigma$ is the occupied MO overlap matrix.
We follow the projection operator form proposed by \citeauthor{Stoll1980} shown in Eq.~\eqref{eq:Stoll_projector}. \cite{Stoll1980}
Diagonalization of the fragment Fock matrix,  $[\textbf{F}]_{\text{XX}}$ of fragment AO dimension, yields MOs which are absolutely localized on that fragment.

The SCFMI wavefunction allows ALMOs on each fragment to polarize in the presence of ALMOs on all other fragments. 
In terms of physical interactions, the SCFMI wavefunction can capture electrostatics, exchange repulsion, dispersion (as permitted by the XC functional), and polarization between fragments as described by the mean-field theory used.
One important interaction \textit{not captured} by the SCFMI wavefunction is the effect of charge transfer among fragments.
One way to capture this charge delocalization effect is through perturbation theory, which works decently well for weakly interacting systems. \cite{Khaliullin2006}
\Insertnew{Another more expensive but systematic method to capture the effects of charge delocalization is the Generalized X-Pol method which treats dimer states in the spirit of non-orthogonal configuration interaction theory. \cite{Gao2010}}
In terms of computational cost, SCFMI removes the diagonalization bottleneck involved in the SCF iterations by requiring only fragment Fock matrix diagonalizations. 
However, there is no cost saving in constructing the Fock matrix for each SCF iteration.

\subsection{EMFT}
The Embedded Mean-Field Theory (EMFT) is a simple embedding scheme for mean-field \insertnew{electronic structure} theories, which partitions the system and the environment at the single-particle basis set level. \cite{Fornace2015}
EMFT uses a more accurate and usually more expensive mean-field theory (referred to as `high level') for the region of interest, while using an approximate and inexpensive mean-field theory (referred to as `low level') for the rest of the system.
The region of interest is usually one in which a chemical transformation takes place (denoted as A), and the surrounding is usually not directly involved in the chemical transformation (denoted as B).
The EMFT energy is obtained by minimizing the EMFT energy function given by Eq.~\eqref{eq::EMFT_energy_expression}.
\begin{align}
    E_{\text{EMFT}} = E_{\text{low}}[\textbf{P}] - E_{\text{low}}[\textbf{P}^{\text{AA}}] + E_{\text{high}}[\textbf{P}^{\text{AA}}] \label{eq::EMFT_energy_expression}
\end{align}
where $E_{\text{low}}[\textbf{P}]$ is the low level energy computed using the full density matrix $\textbf{P}$,  $E_{\text{low}}[\textbf{P}^{\text{AA}}]$ is the low level energy computed using the AA block of the density matrix, and $E_{\text{high}}[\textbf{P}^{\text{AA}}]$ is the high level energy computed using the AA block of the density matrix.
The core idea is to compute interactions in subsystem A, represented by the AA block of the density matrix \textbf{P}\textsuperscript{AA}, using the high-level mean-field theory. 
In the context of mean-field theories, hybrid density functionals may serve as the best choice for the high-level theory, while semi-local functionals would be suitable choices for low level.
For this combination of hybrid density functional in semi-local density functional, the EMFT energy expression in Eq.~\eqref{eq::EMFT_energy_expression} can be written as Eq.~\eqref{eq::EMFT_energy_hybrid_in_local} where `high' and `low' denote the the hybrid density functional and semi-local density functional respectively, and $\alpha$ is the fraction of exact exchange present in the hybrid density functional.
\begin{align}
    E_{\text{EMFT}} &= E_{\text{xc,low}}[\textbf{P}] + \big( E_{\text{xc,high}}[\textbf{P}^{\text{AA}}] - E_{\text{xc,low}}[\textbf{P}^{\text{AA}}] \big) + \alpha E_{\text{K}}[\textbf{P}]  \label{eq::EMFT_energy_hybrid_in_local} \\
    \textbf{F}_{\text{EMFT}} &= \frac{\partial E_{\text{EMFT}}}{\partial \textbf{P}} \\
    &= \textbf{F}_{\text{xc,low}}[\textbf{P}] + \big( \textbf{F}_{\text{xc,high}}[\textbf{P}^{\text{AA}}] - \textbf{F}_{\text{xc,low}}[\textbf{P}^{\text{AA}}] \big) + \alpha \textbf{K}^{\text{AA}}[\textbf{P}]  \label{eq::EMFT_Fock_hybrid_in_local}
\end{align}
The EMFT Fock matrix expression follows directly from the energy expression  (shown in Eq.~\eqref{eq::EMFT_Fock_hybrid_in_local}). 
The EMFT paper \replacewith{mentions}{reports} two models for computing exact exchange, EX0 and EX1 (shown in Eq.~\ref{eq::EMFT_EX0} and Eq.~\ref{eq::EMFT_EX1} respectively).
\begin{align}
    E_{\text{K,EX0}}[\textbf{P}^{\text{AA}}] &= - \frac{1}{4} \sum _{\mu, \nu, \sigma, \lambda \in A} \textbf{P}^{\text{AA}}_{\lambda, \mu} \big( \mu \nu | \sigma \lambda \big) \textbf{P}^{\text{AA}}_{\nu \sigma} \label{eq::EMFT_EX0} \\
     E_{\text{K,EX1}}[\textbf{P}^{\text{AA}}] &= - \frac{1}{4} \sum_{\substack{\nu,\sigma \\ \mu,\lambda \in A }}  \textbf{P}^{\text{AA}}_{\lambda, \mu} \big( \mu \nu | \sigma \lambda \big) \textbf{P}_{\nu \sigma}  \label{eq::EMFT_EX1}
\end{align}
The EX0 requires the construction of only the AA block of the exact exchange matrix (\textbf{K}) from the AA block of the density matrix (\textbf{P}\textsuperscript{AA}), providing substantial savings in cost.
However, this model of computing the exact exchange suffers from severe numerical instability due to unnatural density shift to the part of the system containing exact exchange (A), as shown in Fig.~S1. 
This is caused, in part, due to the lack of balancing counterpart terms in the semi-local functionals to the negative-definite exact exchange contribution of $\alpha E_{\text{EX0}}[\textbf{P}^{\text{AA}}]$ of hybrid DFT to the EMFT energy expression.
In the original EMFT paper, \cite{Fornace2015} this is overcome by using the more expensive EX1 model for exact exchange which computes the exact exchange matrix using the full density matrix, instead of the just the AA block. 
Another technique to alleviate this problem, while still using the the EX0 model, comprises of doing the computations in AO basis orthogonal between A and B. \cite{Ding2017}
This approach in turn breaks the fragment sparsity in \textbf{C} and is therefore not suitable for our purposes.
In this work, we employ the EX0 model for computing exact exchange but constrain the particle number on each fragment by solving for the SCFMI equations instead of the full SCF equations. 
This fixes the numerical instability problem in EMFT.
While the problem of unphysical collapse in the wavefunction is solved, particle number fluctuation (i.e. charge delocalization) is allowed at a two-body level.

\subsection{PolBE theory}

\begin{figure}
    \centering
    \includegraphics[width=\textwidth]{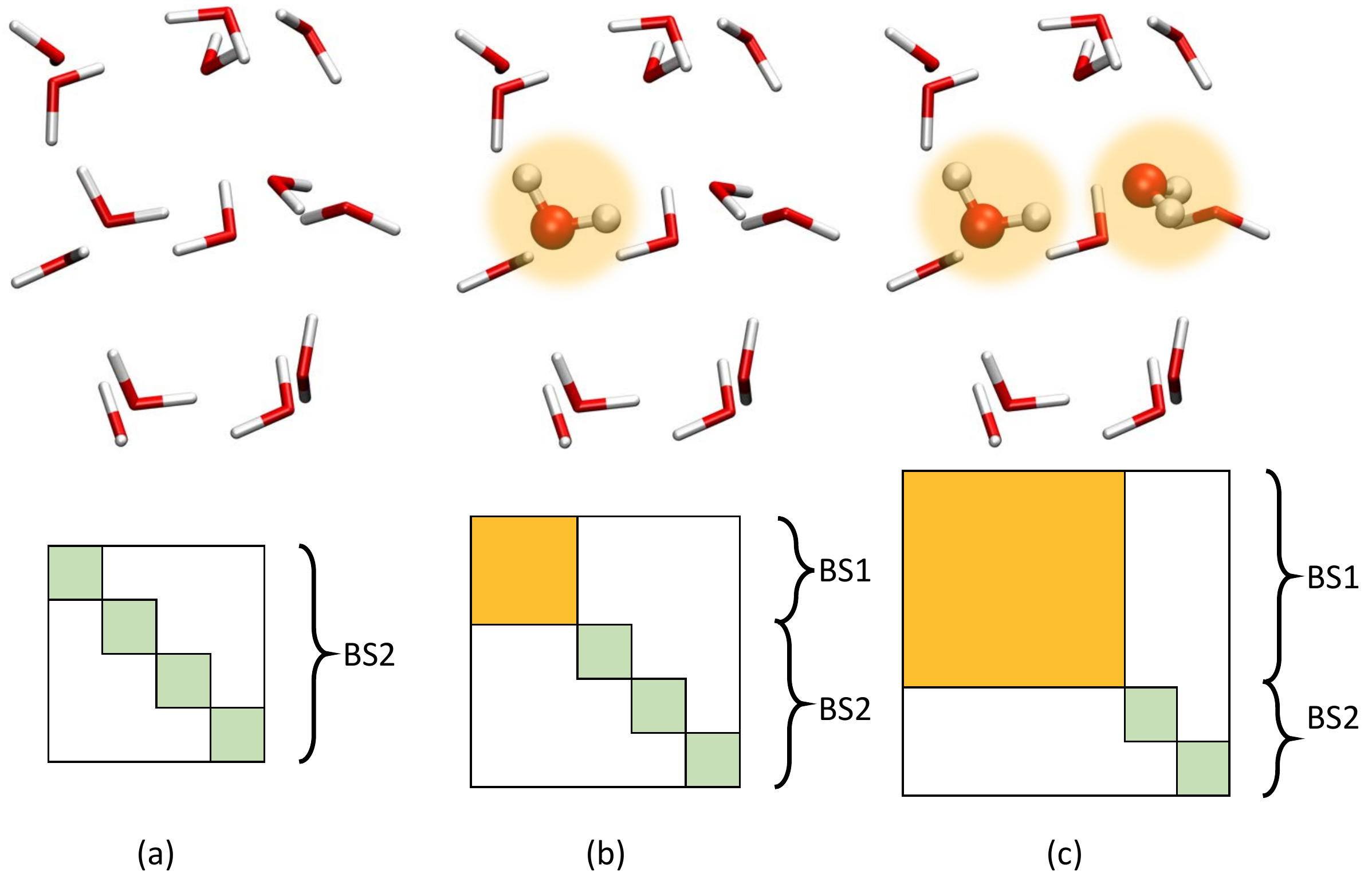}
    \caption{A schematic of the PolBE algorithm. The top panel schematically shows the density functional and basis set used for each of the monomers. Highlighted monomers are treated at XC1/BS1, and the others are treated at XC2/BS2. The bottom panel shows the structure of the MO coefficient matrix and the size of each of the blocks  (a) V\textsubscript{0}: All fragments are treated at XC2/BS2 (b) V\textsubscript{1}: One of the fragments is treated at XC1/BS1 while the others are treated at XC2/BS2 (c) V\textsubscript{2}: Two of the fragments are treated at XC1/BS1 while the others are treated at XC2/BS2}
    \label{fig:PolBE_scheme}
\end{figure}

As mentioned before, we have identified that the SCFMI wavefunction is well-suited to achieve the delicate balance between computational feasibility and incorporating many-body effects into lower order terms.
We use EMFT to treat the system using an accurate, expensive mean-field theory while treating the environment using a more approximate and relatively inexpensive mean-field theory.
Further computational advantages can be achieved by using a smaller basis set for the environment.
For the rest of the paper, we will use XC1 to denote the high level functional and XC2 to denote the low level functional. 
Similarly, BS1 will refer to the larger basis set and BS2 to the smaller basis set.
\begin{enumerate}
    \item The first step in PolBE is to calculate the polarized density of the full system at XC2/BS2.
    The MO coefficient matrix, illustrated in Fig.~\ref{fig:PolBE_scheme}(a), is completely represented in BS2.
    This approximate polarized density is used for embedding all $k$-body terms treated using XC1/BS1. 
    \item For a system consisting of $F$ fragments, each 1-body term is calculated at XC1/BS1 in the presence of the approximate polarized density of the remaining $(F-1)$ fragments.
    Fig.~\ref{fig:PolBE_scheme}(b) shows one of the F 1-body terms in which the monomer is represented in BS1 while the rest of the environment is represented in BS2.
    In each monomer calculation, EMFT with EX0 is used to build Fock matrix treating the monomer at XC1 while the rest of environment is treated at XC2. 
    Instead of diagonalizing the full Fock matrix, SCFMI is used to construct a projected Fock matrix only for the monomer fragment which is diagonalized, and its density is updated until convergence is reached. It should be emphasized that the density of the environment is not updated.
    \item At a 2-body level, dimers are formed by combining all possible monomers into a larger fragment and treated at XC1/BS1 as described earlier.
    One of the $^NC_2$ dimers is shown as a large fragment consisting of two monomers and is represented in BS1 as shown in Fig.~\ref{fig:PolBE_scheme}(c).
\end{enumerate}


\newpage

\section{Computational Details}
$\omega$B97M-V, a combinatorially optimized range-separated hybrid meta-GGA functional with VV10 non-local correlation, \cite{Mardirossian2016} has been known to provide very good performance for a range of properties, including binding energies for non-covalent dimers and clusters. \cite{Mardirossian2017} 
We used this functional to represent the system denoted earlier as XC1. 
We used PBE, a more inexpensive and less accurate GGA functional, to represent the environment (XC2).  \cite{Perdew1996}
Def2-TZVPPD, a triple-zeta basis set with polarization and diffuse functions was used for BS1.\cite{Rappoport2010a}
Def2-SV(P), a double-zeta split-valence basis set with polarization functions only on heavy atoms was used for BS2. \cite{Weigend2005} 
A quadrature grid consisting of 99 Euler-MacLaurin radial points and 590 Lebedev angular points was used for integrating both the XC functionals.
SG-1 was used for the non-local VV10 part.\cite{Gill1993} 
Binding energies reported below are vertical binding energies, computed as the difference between the energy of the full system and the sum of fragment energies at the super system geometry.
The reference SCF binding energy was computed using XC1/BS1 ($\omega$B97M-V/def2-TZVPPD);
the goal of PolBE is to accurately reproduce the full SCF energy of the supersystem.
\Replacewith{PolBE was implemented in a developmental version of Q-Chem 5.0.}{PolBE was implemented in the new SCF manager of Q-Chem\cite{Shao2015} which was developed over Refs.~\citenum{Lee2018, Lee2019, Lee2019a, Bertels2019}.}

\newpage

\section{Results and Discussion}
 
\subsection{Water Trimer PES}
\begin{figure}
     \centering
     \begin{subfigure}[t]{0.29\textwidth}
         \centering
         \includegraphics[width=\textwidth]{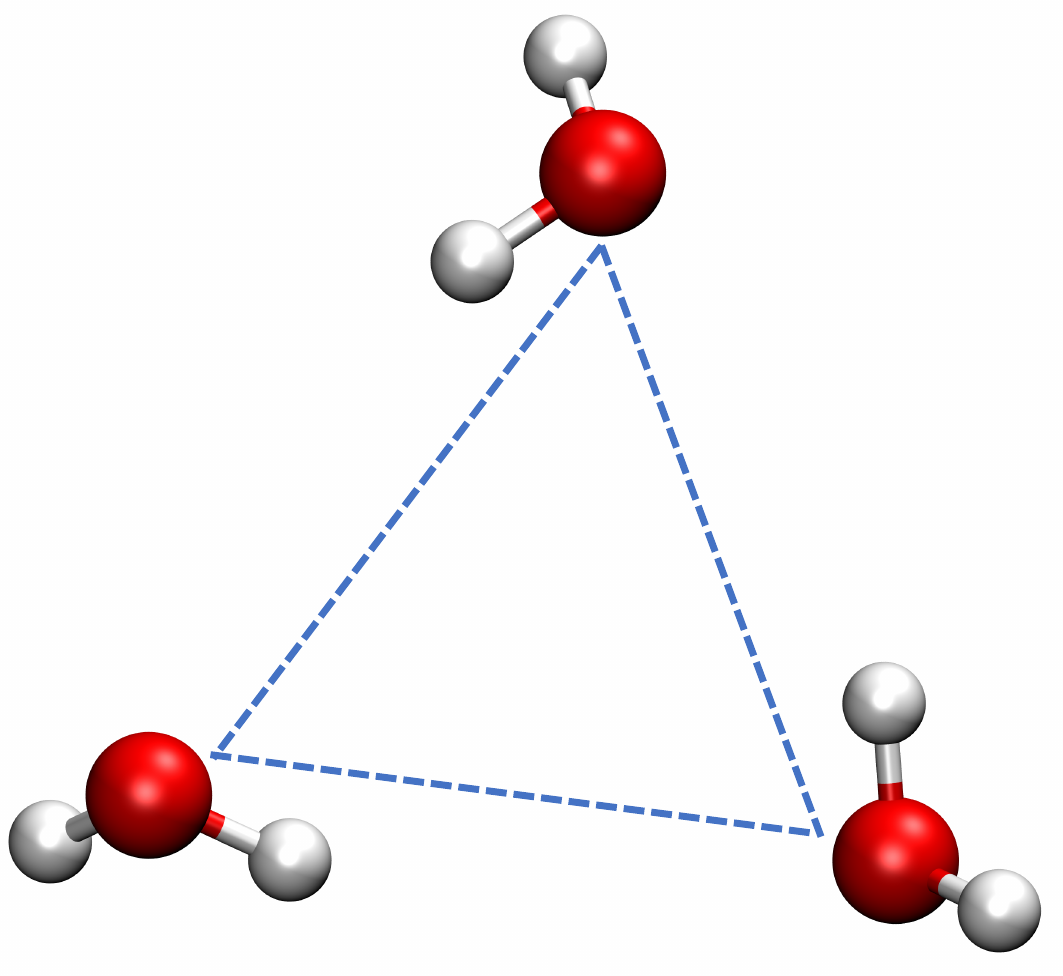}
         \caption{}
     \end{subfigure}
     \begin{subfigure}[t]{0.69\textwidth}
         \centering
         \includegraphics[width=\textwidth]{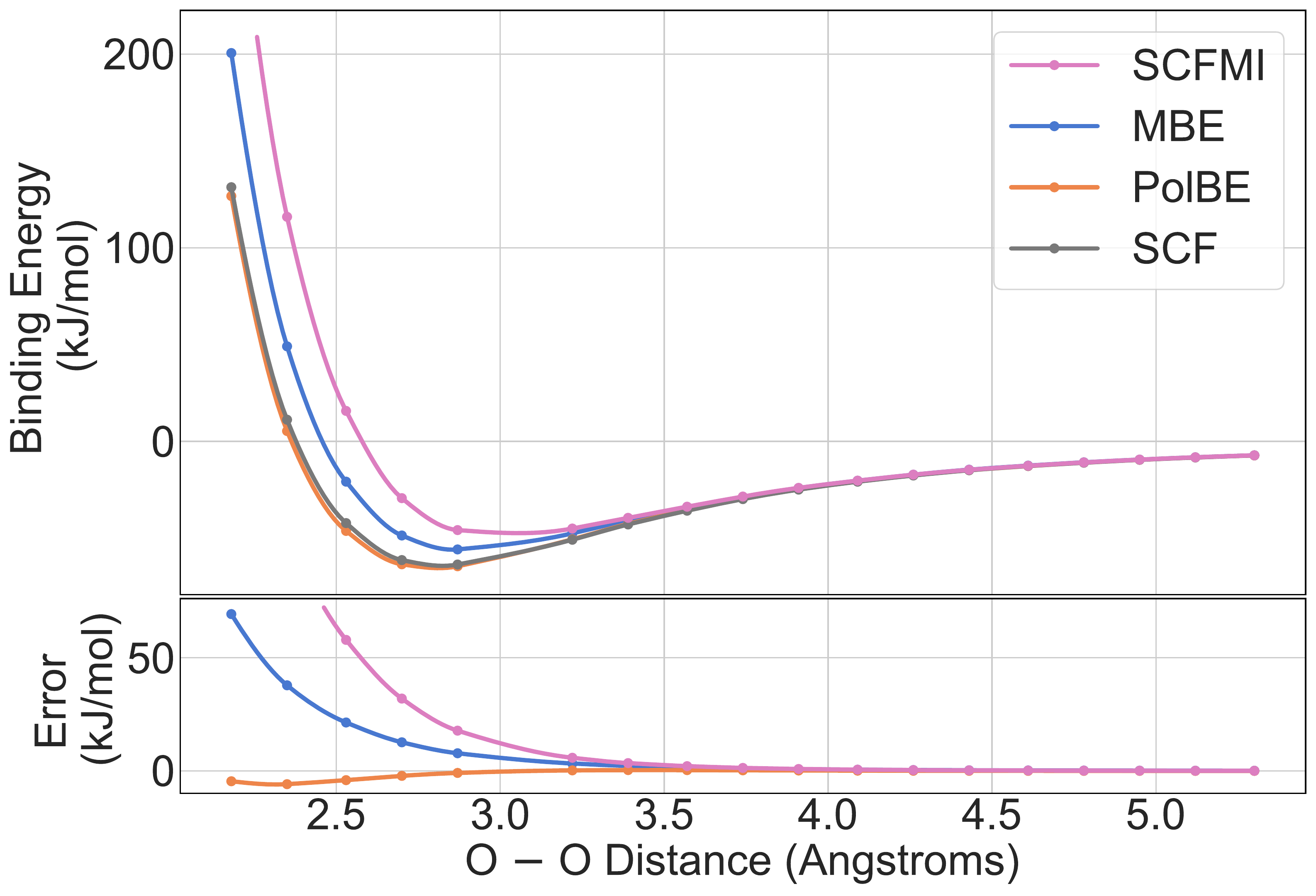}
         \caption{}
     \end{subfigure}
        \caption{(a) Water trimer formed by stretching molecules away from each other (b) Water PES as computed by full SCF, vacuum MBE, and PolBE. Associated errors are shown in the bottom panel.}
        \label{fig:water_trimer}
\end{figure}

A desirable property of an embedded Many-Body expansion is that it should provide a smooth Potential Energy Surface (PES) and consequently a continuous nuclear gradient so that it can used for performing geometry optimizations, transition state searches, and molecular dynamics simulations. \cite{Mata2006, Subotnik2008}
In order to evaluate the performance of PolBE \replacewith{with a smoothly varying}{as a function of} nuclear \replacewith{coordinate}{position}, we have investigated the PES of water trimer.
PolBE relies on natural fragmentation arising from the lack of covalent bonding between different molecules in the cluster.
There are no cut-off parameters which depend on real-space distances, and thus PolBE provides smooth potentials as shown in Fig.~\ref{fig:water_trimer}. 
PolBE is also trivially size-consistent as it recovers the non-interacting limit at large distances.
In the interacting regime, PolBE binding energies are in good agreement with the SCF reference binding energies. 
PolBE also successfully recovers the repulsive part of the PES, while the EX0 model of EMFT cannot as shown in Fig.~S1.

This numerical instability does not appear in PolBE despite using the inexpensive EX0 model for exact exchange as the ALMO constraint prevents any charge transfer between fragments.
However, it should be emphasized that the lack of charge transfer between fragments simply yields inadequate binding energies (as shown in the SCFMI curve in Fig.~\ref{fig:water_trimer}(b)).
Therefore, allowing for pairwise charge transfer as is done in PolBE is very effective in recovering the missing charge transfer interaction energy.
\Insertnew{Non-additive dispersion interactions are also not captured by PolBE.
These higher-order interactions are typically very small.
In the case of water trimer at equilibrium geometry, higher order dispersion interaction would contribute an additional 0.06 kJ/mol to the PolBE binding energy.}
\subsection{Water pentamer conformers}

\begin{table}[]
\caption{\Replacewith{Binding energies and relative binding energies and associated errors for the SCF reference, vacuum MBE, and PolBE.}{Binding energy errors for the cage and cyclic conformers and relative binding energy ($\Delta E_\text{bind} = E_\text{bind}\text{(cage)} $--$€" E_\text{bind}\text{(cyclic)}$) of the cage conformer with respect to the cyclic conformer as predicted by vacuum MBE truncated at 2-body (MBE-2), vacuum MBE truncated at 3-body (MBE-3), and PolBE truncated at 2-body (PolBE-2). The SCF binding energies for the cage and cyclic conformers are $-153.7$ and $-160.7$ kJ/mol.} All values are in kJ/mol.
}
\label{tab:water_pentamer_conformers}
\begin{tabular}{c|rr|r}
\cline{2-3}
                              & \multicolumn{2}{c|}{E\textsubscript{bind} error}                         &                                                \\ \cline{2-4} 
                              & \multicolumn{1}{c|}{cage} & \multicolumn{1}{c|}{cyclic} & \multicolumn{1}{c|}{$\Delta$ E\textsubscript{bind}} \\ \hline
\multicolumn{1}{|c|}{MBE-2}   & $28.4$                      & $45.4$                        & \multicolumn{1}{r|}{$-9.9$}                                          \\ \cline{1-1}
\multicolumn{1}{|c|}{MBE-3}   & $1.1$                       & $5.9$                         & \multicolumn{1}{r|}{$2.3$}                                              \\ \cline{1-1}
\multicolumn{1}{|c|}{PolBE-2} & $-7.2$                      & $-4.5$                        & \multicolumn{1}{r|}{$4.3$}                                              \\ \hline
\end{tabular}
\end{table}

\begin{figure}
     \centering
     \begin{subfigure}[b]{0.49\textwidth}
         \centering
         \includegraphics[width=\textwidth]{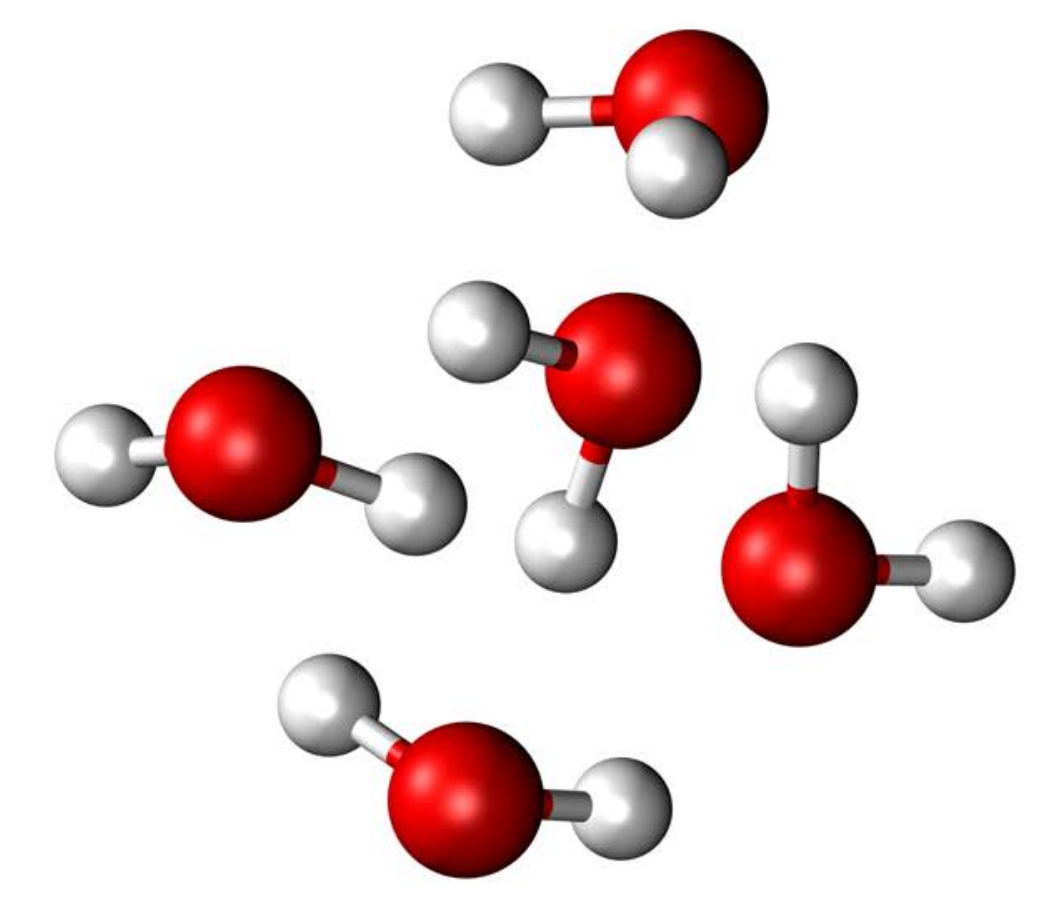}
         \caption{}
         \label{fig:water_pentamer_cage}
     \end{subfigure}
     \begin{subfigure}[b]{0.49\textwidth}
         \centering
         \includegraphics[width=\textwidth]{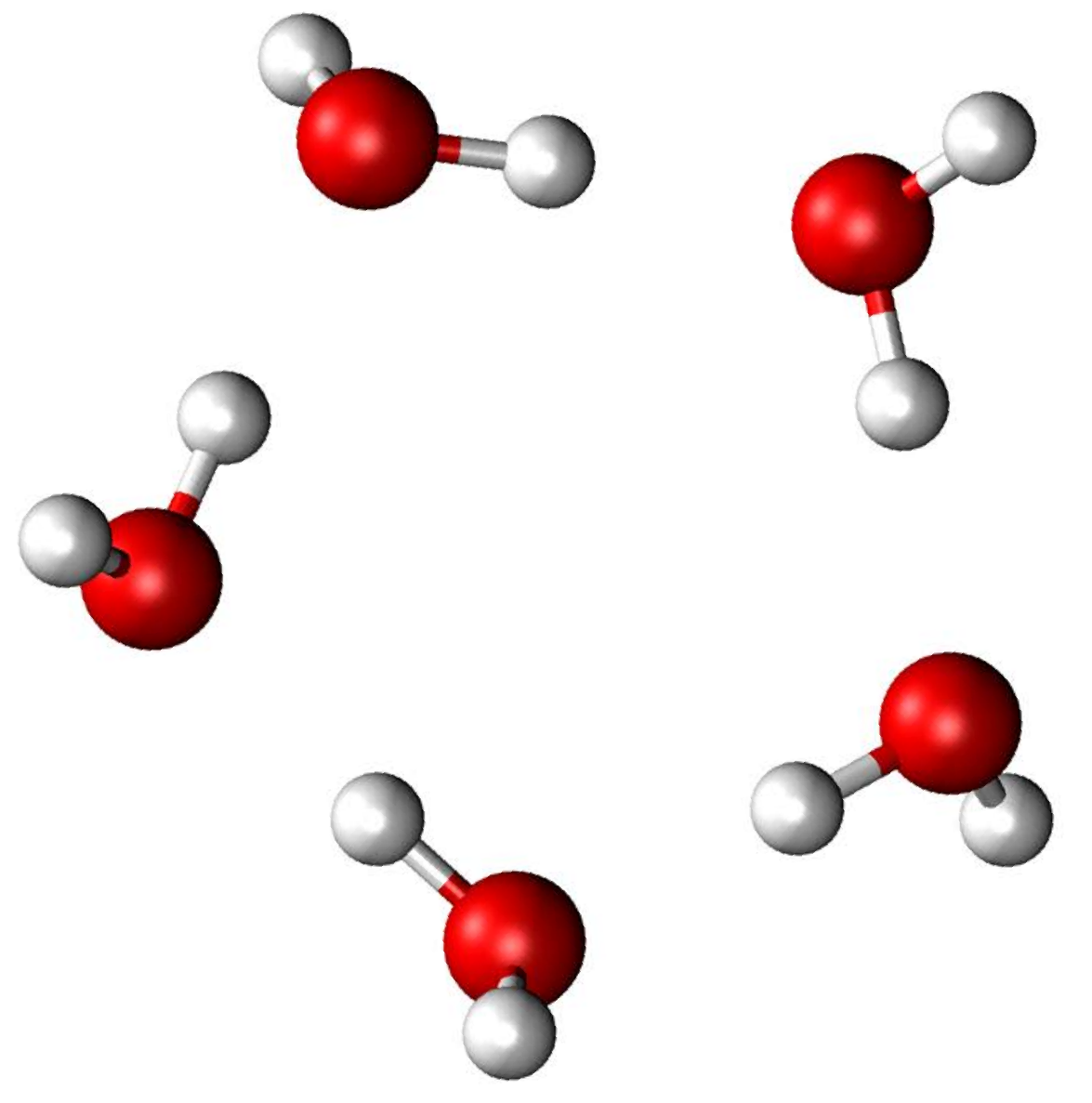}
         \caption{}
         \label{fig:water_pentamer_cyclic}
     \end{subfigure}
        \caption{(a) Cage conformer of water pentamer (b) Cyclic conformer of water pentamer.}
        \label{fig:water_pentamer}
\end{figure}

When a many-body expansion is truncated at the 2-body level,higher order binding cooperativity is not captured.
The change in the binding affinity of molecule upon being bound to another water molecule is termed cooperativity.
The phenomenon of cooperativity is found ubiquitously in chemical systems including water, \cite{Xantheas2000} methanol,\cite{Sum2000} and hydrogen fluoride. \cite{Rincon2001}
The importance of understanding cooperativity cannot be overemphasized as it is crucial for understanding the bulk phase properties of these molecules.
For instance, it enhances the strength of hydrogen bonding in water at a molecular level leading to anomalous adsorption properties in bulk water. \cite{Luck1998}
Cooperativity or non-pairwise additivity can be quantified as the amount of interaction energy that is not captured by the many-body expansion truncated at the 2-body level.
In this work, we show how the higher order effect of cooperativity may be captured at a 2-body level using the PolBE.

There has been considerable interest in predicting the relative energies of various molecular clusters,  both qualitatively and quantitatively. \cite{Yuan2017, Temelso2015, Friedrich2016}
More often than not, hydrogen bonded molecules in clusters interact in a non-pairwise fashion,\cite{Sum2000,Rincon2001} making the total binding energy not a simple sum of pairwise interaction energies.
It has been shown that homodromic hydrogen bonding ring-like networks are associated with the highest cooperativity. \cite{Xantheas2000}
This defining characteristic of these structures make them the most stable conformer in small water clusters, from trimer through pentamer, even though they contain only one hydrogen bond per water molecule while other conformers have more than one hydrogen bond per water molecule.

In this section, we investigate the case of two conformers of water pentamer: cyclic and cage using geometries taken from Ref.~\citenum{Temelso2011}.
The cyclic conformer, with the homodromic ring-like structure, exhibits more cooperativity in binding than the cage conformer.
Vacuum MBE trucated at the 2-body level makes errors of 28.4 and 45.4 kJmol\textsuperscript{-1} for the cage and cyclic conformer respectively showing that that the cyclic conformer has more cooperativity than the cage variant.
Vacuum MBE not only predicts quantitatively wrong binding energies, but also incorrectly predicts that the cage conformer is the more stable than the cyclic one.
Vacuum MBE needs the 3-body term in order to capture this many-body effect, as shown in Table.~\ref{tab:water_pentamer_conformers}.
A full DFT calculation ($\omega$B97M-V/def2-TZVPPD) on these conformers shows that the cyclic conformer is more stable than the cage by 7.0 kJ/mol.
PolBE which correctly embeds the 1-body and 2-body terms in an approximate representation of the environment makes much smaller binding energy errors of 7.2 and 4.5 kJ/mol.
PolBE is able to correctly pre-polarize the monomers, thus decomposing the interactions in a pairwise additive fashion.
This demonstrates the ability of our 0\textsuperscript{th} order wavefunction (SCFMI wavefunction) to precisely capture the mutual polarizations between the different water molecules which is clearly a many-body effect.
PolBE also predicts the relative energetics correctly, predicting that the cyclic conformer is more stable with a relative energy error of 2.7 kJ/mol, smaller than chemical accuracy of 1 kcal/mol.
In the context of molecular simulation, it is more important to get the relative binding energy ordering and separations correctly rather than absolute binding energies. 
While vacuum MBE needs to be carried out till the third order to even get the relative ordering of conformers correct qualitatively, PolBE can achieve both the qualitative and quantitative relative energy ordering correct at the two-body level.

The water simulation community has recognized the importance of these higher-body effects and has tried to incorporate them either implicitly through inclusion of classical polarization \cite{Ponder2010,Ren2003} or explicitly by including 3-body and other higher order terms. \cite{Tainter2011,Babin2013a, Cisneros2016, Medders2015, Reddy2016}
This parametrization is further complicated by the fact that different phases of water (clusters, liquid water, and ice) show different magnitudes of non-additivities. \cite{Dang1997,Burnham1999}
PolBE does not contain any parametrization, and \Replacewith{can be expected to work as well as the `high level' density functional.}{can approximate the `high level' density functional with reasonable accuracy.} 
PolBE also suggests an alternative way to account for these many-body effects, through the use of an inexpensive, albeit crude, zeroth order wavefunction.

\subsection{\ce{CO2} clusters}

\begin{figure}
     \centering
     \begin{subfigure}[b]{0.49\textwidth}
         \centering
         \includegraphics[width=\textwidth]{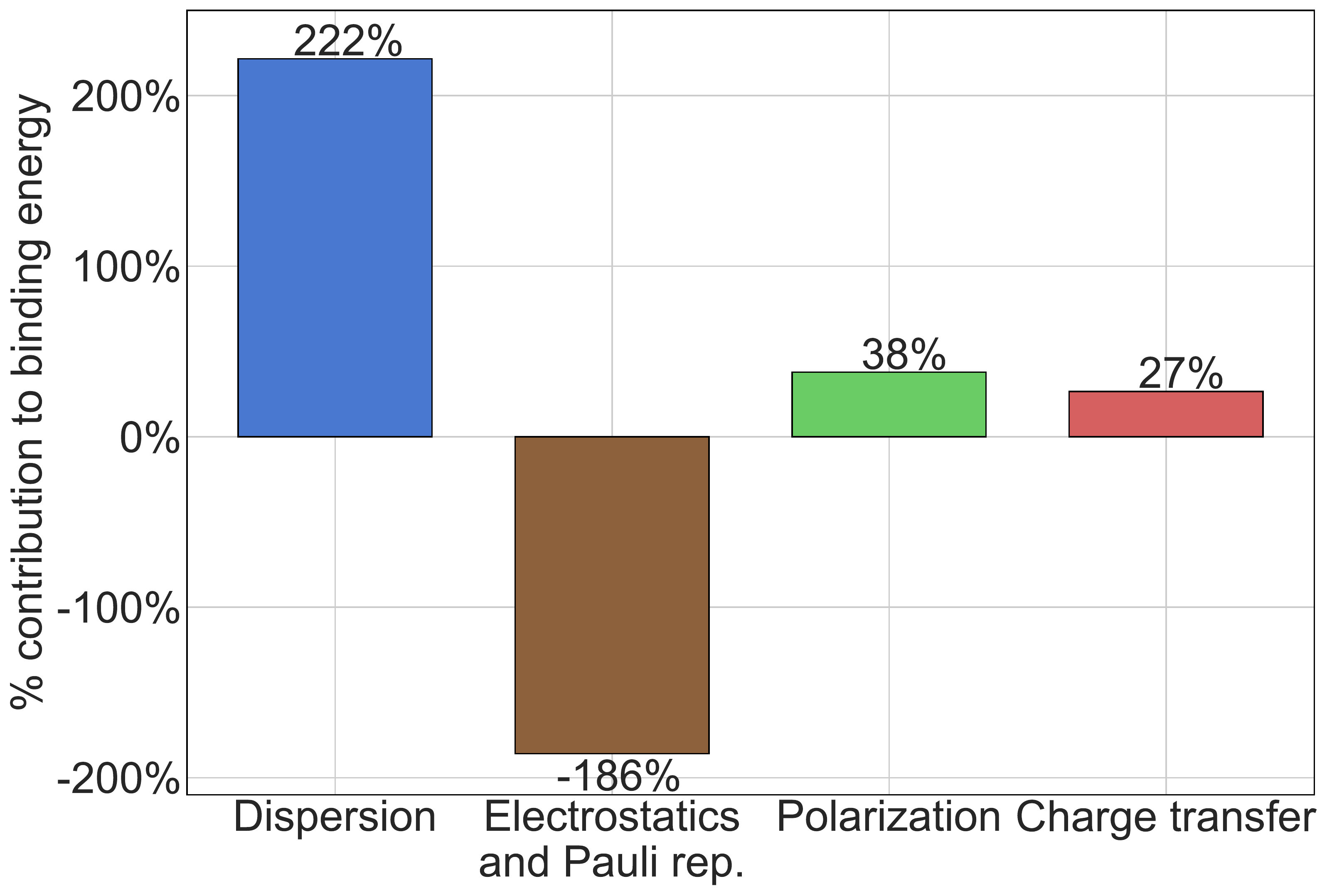}
         \caption{}
         \label{fig:CO2_EDA}
     \end{subfigure}
     \begin{subfigure}[b]{0.49\textwidth}
         \centering
         \includegraphics[width=\textwidth]{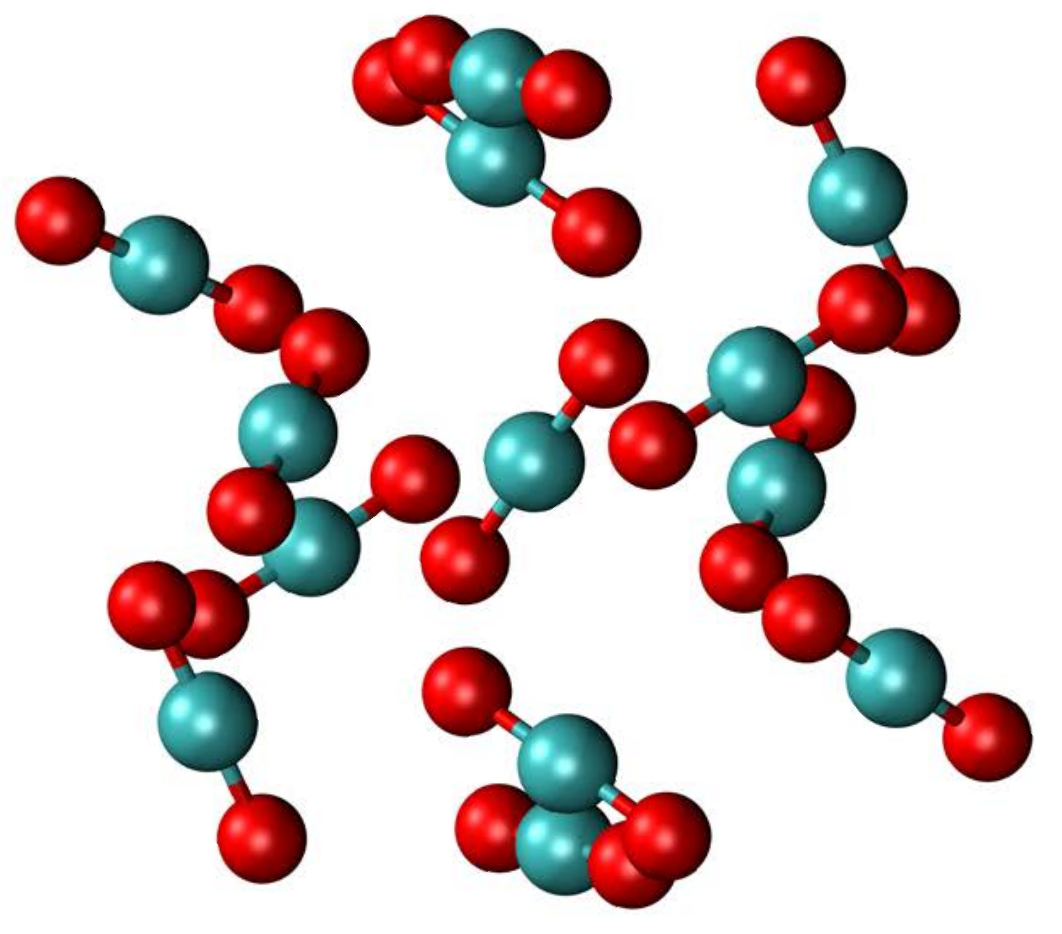}
         \caption{}
         \label{fig:CO2_13mer}
     \end{subfigure}
     \begin{subfigure}[b]{0.49\textwidth}
         \centering
         \includegraphics[width=\textwidth]{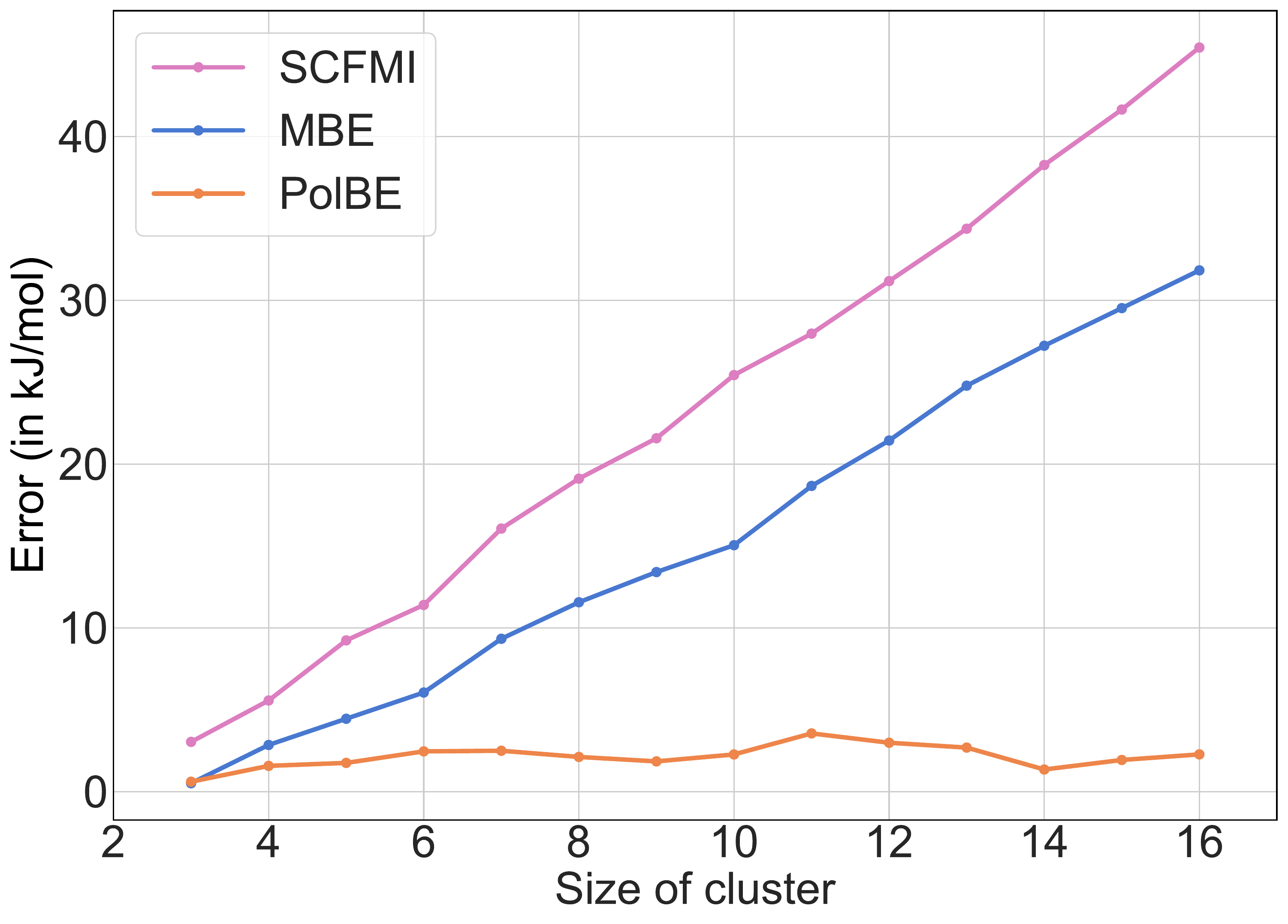}
         \caption{}
         \label{fig:CO2_total}
     \end{subfigure}
     \begin{subfigure}[b]{0.49\textwidth}
         \centering
         \includegraphics[width=\textwidth]{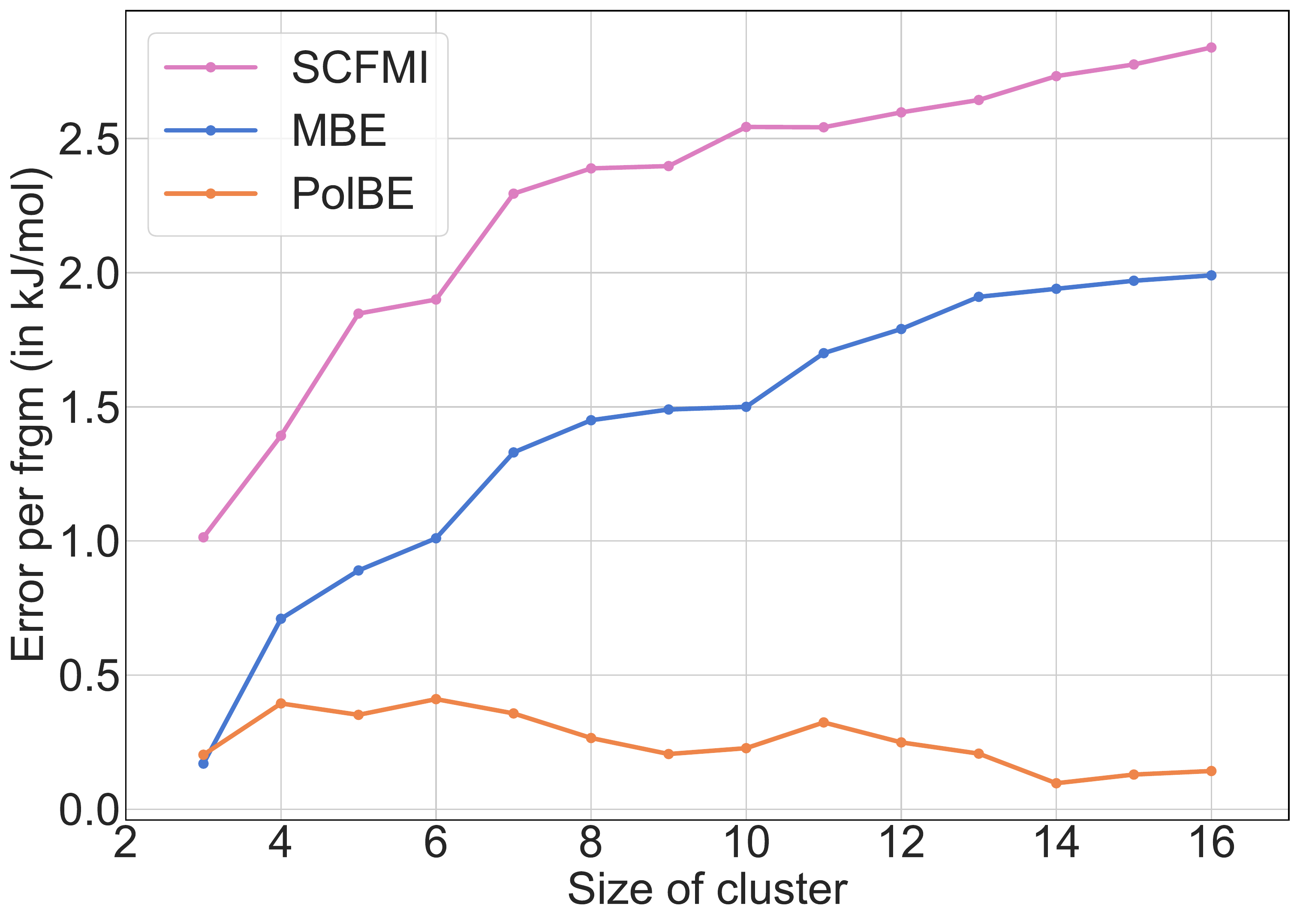}
         \caption{}
         \label{fig:CO2_perfrgm}
     \end{subfigure}
        \caption{(a) Energy decomposition analysis for all \ce{CO2} clusters considered in this study. \insertnew{Dispersion is the interaction between induced dipoles of different fragments. Pauli repulsion is the steric energetic penalty that occurs when two non-bonded atoms are forced to occupy the same space. Electrostatics is the energy associated with the coulombic interactions between fragments. Polarization is the energy associated with relaxation of molecular orbitals. Charge transfer is the energy associated with the flow of electrons between the fragments.} (b) A \ce{CO2} cluster containing 13 molecules (c) SCFMI, vacuum MBE, and PolBE errors as function of \ce{CO2} cluster size (d) SCFMI, vacuum MBE, and PolBE errors per fragment as function of \ce{CO2} cluster size. For comparison, the reference binding energy for 16-mer is 225.3 kJ/mol.}
        \label{fig:CO2_graphs}
\end{figure}

There has been widespread interest in studying \ce{CO2} molecular clusters because of their presence in \replacewith{Earth's}{earth's} atmosphere and that of other planets. \cite{Wang2009}
There is a significant interest in the green chemistry community as supercritical \ce{CO2} can be used as a benign solvent on an industrial scale. \cite{Desimone2002}
Theoretical studies on \ce{CO2} clusters have served as a bridge to understand its condensed phase properties.

\ce{CO2} are classic examples of weakly-bound clusters.
A study on the intermolecular potential of \ce{CO2} has shown that dispersion is one of the leading terms governing the interaction. \cite{Bukowski1999}
We performed an ALMO-based energy decomposition analysis \cite{Khaliullin2007, Horn2015,Horn2016} of the binding energy of a single \ce{CO2} molecule completely surrounded by other \ce{CO2} molecules for all the clusters considered in this study.
This decomposition\insertnew{, as summarized in Fig.~\ref{fig:CO2_graphs}(a),} shows that polarization, which is the energy associated with the relaxation of molecular orbitals on each fragment in response to molecular orbitals on other fragments, contributes 38\% of the interaction energy with charge transfer causing an additional stabilization of 27\%.
\insertnew{Nevertheless, as shown in Fig.~\ref{fig:CO2_graphs}(a)}, dispersion is the main mechanism which binds these \ce{CO2} clusters, and the ability to describe it accurately is critical to understanding the molecular processes governing the chemistry of carbon dioxide.
In fact, the semi-local functional PBE used for representing the environment, which does not contain any dispersion corrections, predicts that the \ce{CO2} clusters are unbound.
The VV10 non-local correlation functional is a sophisticated method to accurately capture the effect of dispersion.
The high level functional used in this study, $\omega$B97M-V, contains the VV10 non-local correlation functional and correctly predicts the interaction energy of \ce{CO2} trimer to be -14.98 kJ/mol (CCSD(T)\bibnote{The CCSD(T) interaction energy reported in this work have been extrapolated to the Complete Basis Set (CBS) limit using the focal point approximation analysis. $E_{CCSD(T)/CBS} \approx E_{HF/aug-cc-pV5Z}^{SCF} + E_{MP2/aug-cc-pVQZ \rightarrow aug-cc-pV5Z}^{corr} + (E_{CCSD(T)/aug-cc-pVTZ}^{corr} - E_{MP2/aug-cc-pVTZ}^{corr}) $. The MP2 correlation energy was extrapolated to uses the 2-point extrapolation formula \cite{Helgaker1997} using the the aug-cc-pVQZ and the aug-cc-pV5Z basis sets. } interaction energy is -15.96 kJ/mol). 
This shows that the correct treatment of dispersion is very much essential in predicting binding energies of such systems accurately.

We have investigated the ability of PolBE to predict interaction binding energies for a series of \ce{CO2} clusters with geometries taken from Ref. \citenum{Lemke2013}.
In Ref. \citenum{Lemke2013}, the geometries for the \ce{CO2} clusters were optimized at M06/aug-cc-pVDZ. 
The binding energy per monomer increases with system size from $-5.1$ kJ/mol for a trimer to $-14.1$ kJ/mol for the 16-mer.
The binding energy per monomer shows signs of saturation starting at the 13-mer, as it only increases by 0.5 kJ/mol from 13-mer to 16-mer.
The saturation of binding energy per monomer at such a small size of cluster is a consequence of two effects: 1. the loose packing of \ce{CO2} molecules in the cluster and 2. Dispersion interactions which dominate the binding energy of \ce{CO2} are short-range effects.
The errors in the predicted binding energies for SCFMI (XC1/BS1), vacuum MBE and PolBE have been plotted in Figure \ref{fig:CO2_graphs}(c) as a function of cluster size.
The binding energy error predicted by SCFMI and vacuum MBE truncated at the 2-body level increases continuously with system size.
At all fragment sizes, SCFMI, vacuum MBE, and PolBE underestimate the binding energy but the magnitude of underestimation varies significantly.
The binding energy error prediction of the SCFMI wavefunction is the largest, followed by that of vacuum MBE and then PolBE.
For the 16-mer, the largest \ce{CO2} cluster considered in this study, the vacuum MBE error is 31.83 kJ/mol while the PolBE error is only 2.28 kJ/mol, smaller by more than an order of magnitude.

The error per monomer saturates for all the three methods as shown in Figure \ref{fig:CO2_graphs}(d).
For the largest clusters investigated in this study, the binding energy error per monomer saturates at around 0.15 kJ/mol for PolBE whereas it saturates at 2 kJ/mol for vacuum MBE and 2.8 kJ/mol for SCFMI.
This saturation of error per monomer is very encouraging and its magnitude is expected to be transferable to large condensed phase systems of \ce{CO2}.

\subsection{Water clusters}

\begin{figure}
     \centering
     \begin{subfigure}[b]{0.49\textwidth}
         \centering
         \includegraphics[width=\textwidth]{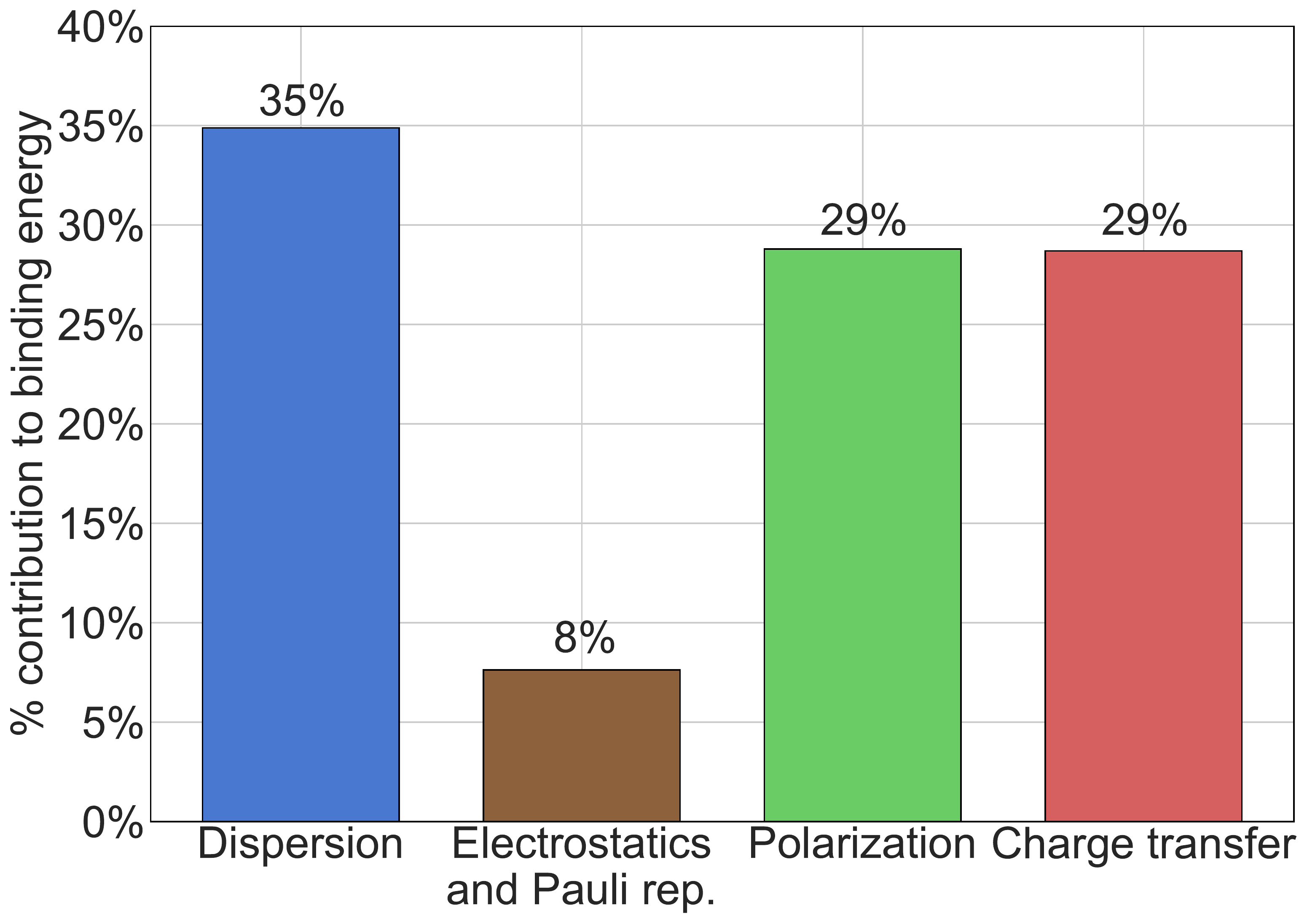}
         \caption{}
         \label{fig:water_EDA}
     \end{subfigure}
     \begin{subfigure}[b]{0.49\textwidth}
         \centering
         \includegraphics[width=\textwidth]{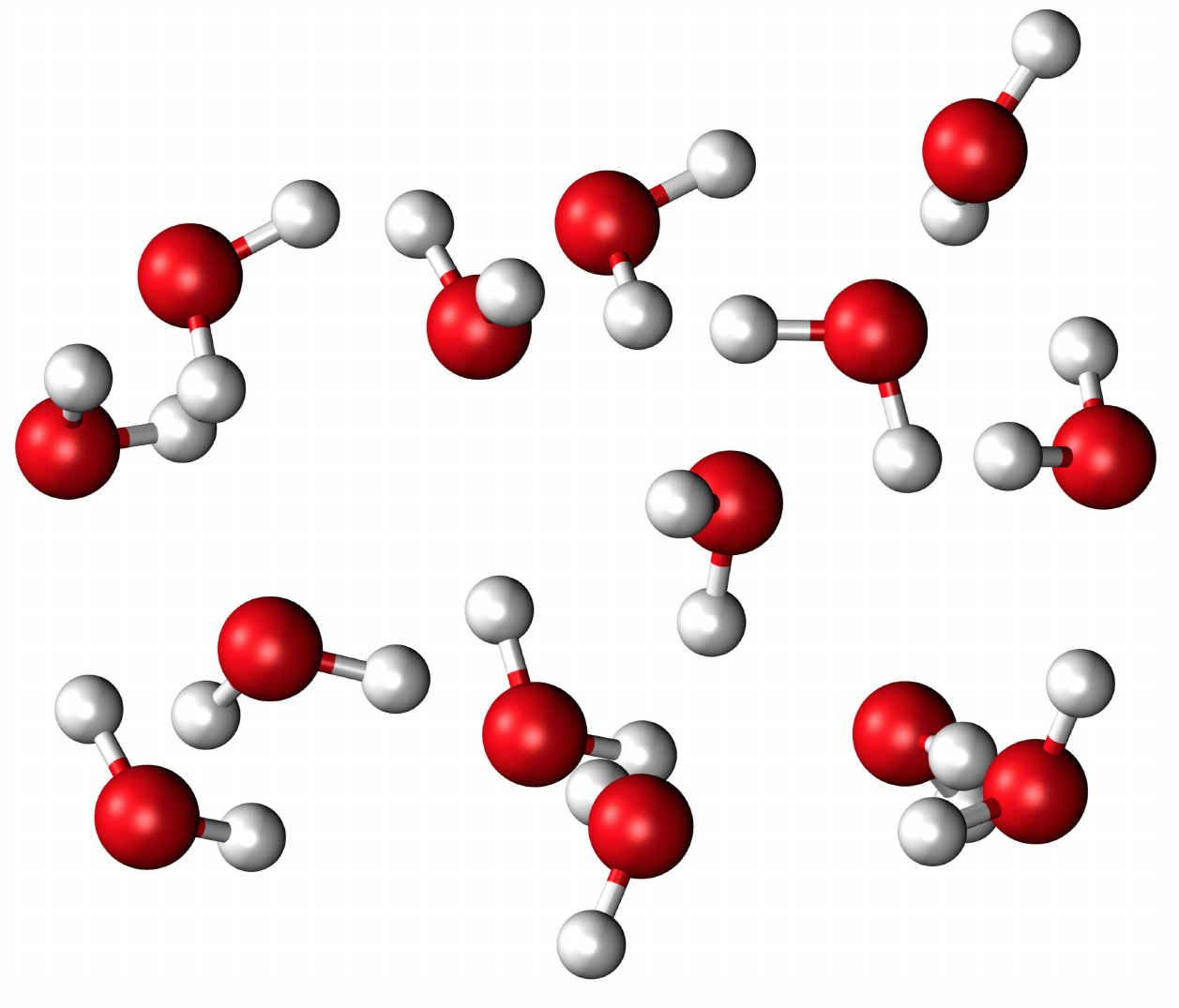}
         \caption{}
         \label{fig:water_21mer}
     \end{subfigure}
     \begin{subfigure}[b]{0.49\textwidth}
         \centering
         \includegraphics[width=\textwidth]{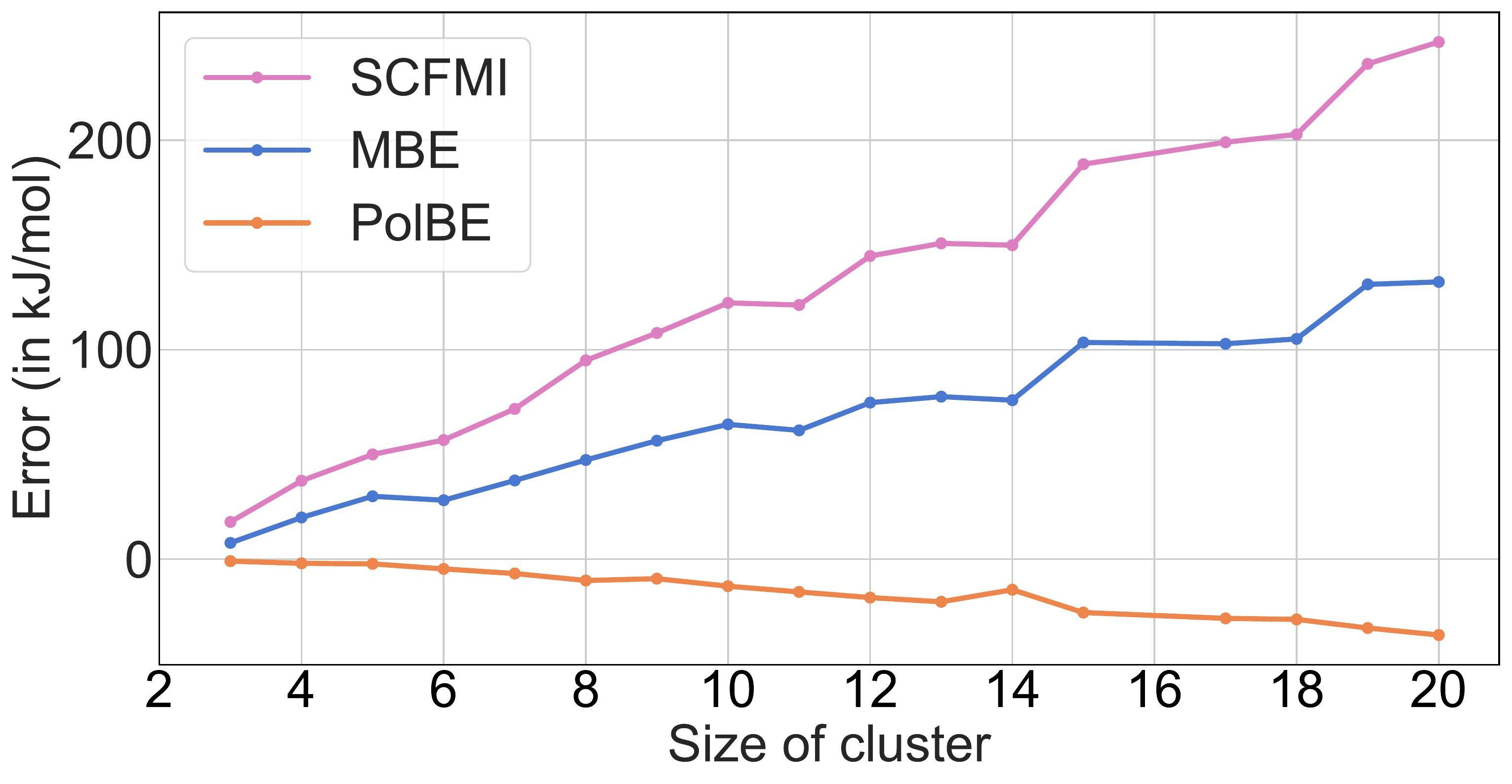}
         \caption{}
         \label{fig:water_total}
     \end{subfigure}
     \begin{subfigure}[b]{0.49\textwidth}
         \centering
         \includegraphics[width=\textwidth]{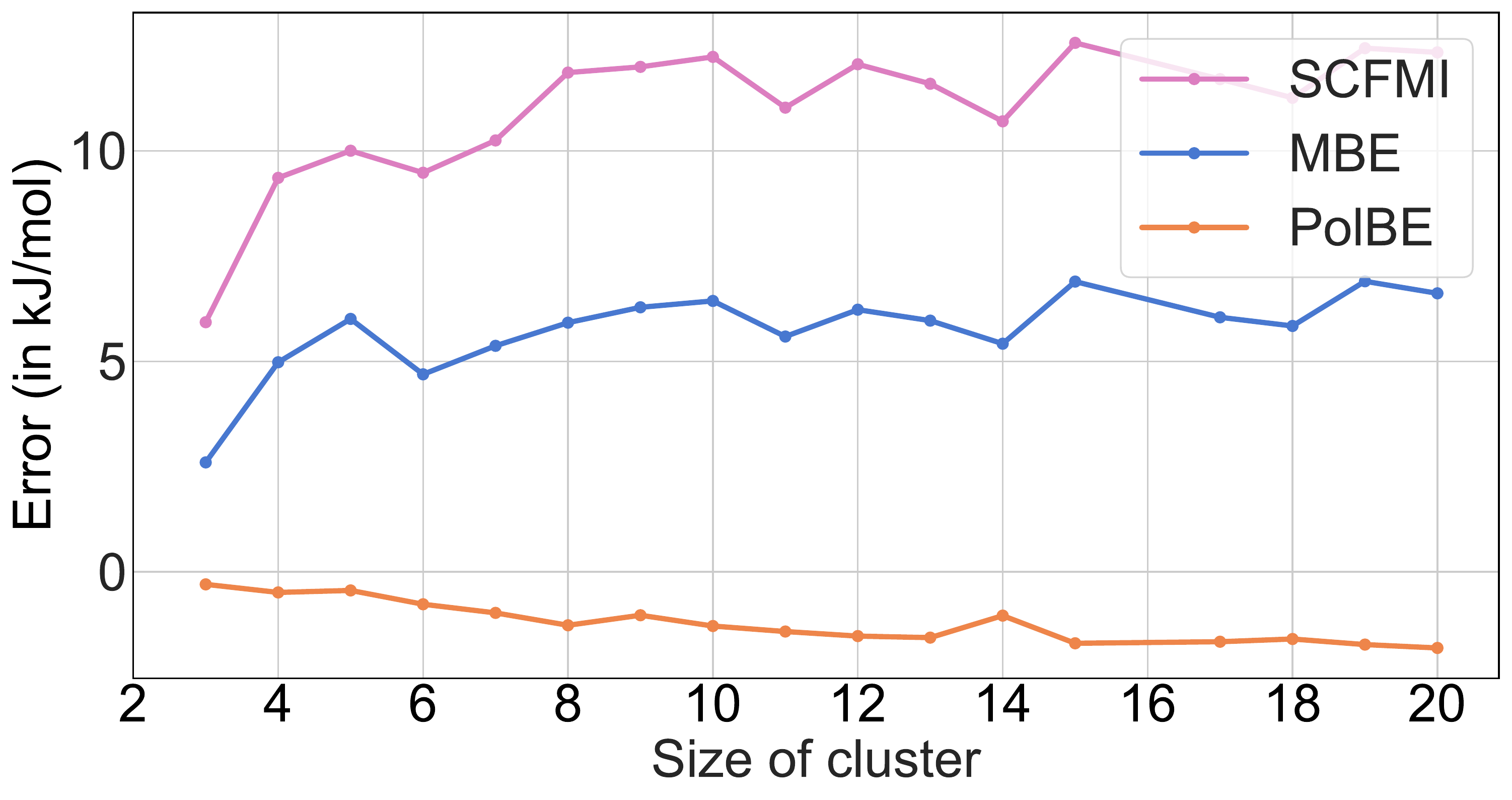}
         \caption{}
         \label{fig:water_perfrgm}
     \end{subfigure}
        \caption{(a) Energy decomposition analysis for all water clusters considered in this study (b) A water cluster containing 14 water molecules (c) SCFMI, vacuum MBE, and PolBE errors as function of water cluster size (d) SCFMI, vacuum MBE, and PolBE errors per fragment as function of water cluster size. For comparison, the total reference binding energy for the water 20-mer is 845.2 kJ/mol.}
        \label{fig:water_graphs}
\end{figure}

Water clusters are of fundamental importance in biological systems and processes and in atmospheric chemistry.\cite{Jordan2017a}
Unlike bulk water, water clusters contain molecules at the surface with unsaturated hydrogen-bonds serving as models for liquid-vapor interfaces.
Almost every possible experimental tool has been used to characterize the structure of water clusters in an attempt to understand its anomalous properties.\cite{Ludwig2001}
It is also possible to use highly accurate ab initio quantum chemical methods to study small water clusters, enabling direct comparison between experiment and theory.

Water molecules interact with each other through hydrogen bonding, participating both as donor and acceptor. 
Water molecules in both water clusters and bulk water interact with each other in a highly non-pairwise additive fashion. \cite{Xantheas2000}
Dispersion is an important mechanism contributing 35\% to the total interaction energy of water clusters \insertnew{as shown in Fig.~\ref{fig:water_graphs}(a)}.
The gas phase dipole moment of 1.855D of water is enhanced in the condensed phase by about 40\%. \cite{Gregory1997,Badyal2000,Batista1998,Silvestrelli1999}
This enhancement in dipole moment coupled with the increasing polarizability of water clusters with size of the cluster \cite{Ghanty2003} leads to stronger binding through magnified polarization contributing about 29\% to the interaction energy.
While different energy decomposition schemes differ on the actual magnitude of charge transfer in a hydrogen bond in water, all methods agree on that fact that charge transfer plays a significant role in hydrogen bonding placing its relative importance somewhere between 20\% to 30\%. \cite{Khaliullin2007,Khaliullin2009}
Performing an ALMO-based energy decomposition of the interaction energy for each term of the many-body expansion has shown that polarization and charge transfer play an important role in the 2-body terms, while polarization dominates the 3-body terms. \cite{Cobar2012}
Charge transfer is mostly a pairwise additive effect and higher order terms are negligible.

In this section, we investigate the binding interactions of water clusters using geometries taken directly from a paper by Gadre and co-workers.\cite{Maheshwary2001}
Binding in water clusters is much stronger than that in \ce{CO2} clusters as can be seen from the larger interaction energies at every cluster size.
The binding energy of the water trimer considered in this study has an interaction energy of 63.6 kJ/mol while the \ce{CO2} trimer has a much smaller interaction energy of 15.4 kJ/mol.
The presence of long-range interactions like dipole-dipole interactions slows down the saturation of binding energy per monomer as well.
Even for the 20 water molecule cluster, the binding energy per water molecule is 42.3 kJ/mol and does not seem to be saturated.
As a consequence, SCFMI, vacuum MBE, and PolBE errors per fragment also show slower signs of saturation.
As in the case of \ce{CO2} clusters, SCFMI and vacuum MBE systematically underbinds water clusters.
The magnitude of the vacuum MBE errors, i.e. the errors arising from truncating the many-body expansion at the 2-body level, is an indicator of the importance of 3-body and higher body effects in the binding of water clusters. 
For a water 20-mer, vacuum MBE underestimates the binding energy by 132 kJ/mol which is about 16 \% of the total interaction energy.
PolBE errors are much smaller, increasing from less than 1 kJ/mol for water trimer to $-$36.2 kJ/mol for the water 20-mer.
It is interesting to note that PolBE consistently overbinds water clusters while it underbinds \ce{CO2} clusters.

The error per fragment of vacuum MBE increases, though not monotonically, from the trimer value of 2.6 kJ/mol to 6.6 kJ/mol for the 20-mer.
SCFMI errors per fragment are much larger, consistently upwards of 11 kJ/mol for large clusters.
PolBE makes much smaller errors per fragment starting at 0.3 kJ/mol for trimer and increasing to 1.8 kJ/mol for the 20-mer.
The underlying many-body effect in water clusters, the polarization term, is treated in a complete many-body sense in PolBE.
Furthermore, charge transfer is predominantly pairwise additive in nature and is treated upto 2-body in PolBE.
The success of PolBE can be attributed to treating these underlying physical interactions accurately and consistently.
Both vacuum MBE and PolBE \insertnew{error} curves show\remove{considerable} kinks \insertnew{as a function of cluster size} (\insertnew{see} Fig.~\ref{fig:water_graphs}(d)) due to the lack of a systematic structure in water clusters with increasing size of the cluster.
Certain geometries of water clusters are less compact than others, leading to \replacewith{smaller number of}{fewer} hydrogen bonds and consequently a smaller interaction energy in comparison to its neighboring clusters.
The kinks\remove{in graphs} closely follow the discontinuities in the binding energy per fragment plot as a function of water cluster size.
Unlike the case of \ce{CO2} where\remove{the} both the binding energy per fragment and the errors per fragment are saturated at the largest \insertnew{cluster} size\remove{of the cluster} considered\remove{in this work}, these quantities continue to increase even at \insertnew{the} 20-mer for water.
Long-range electrostatics are, in part, responsible for this behavior and pose a considerable challenge to simulation of large systems of water.
\Insertnew{In order to investigate the saturation of SCF reference binding energy per fragment and PolBE binding energy error per fragment, we have investigated much larger water clusters $n=32$ and $n=64$.
The SCF binding energy for $n=32$ and $n=64$ is $-26.36$ and $-30.15$ kJ/mol, and are not saturated.
The PolBE binding energy error per fragment is $-3.4$ and $-3.9$ kJ/mol for $n=32$ and $n=64$ respectively.
We see that the the PolBE binding energy error per fragment is saturating with cluster size, and hence showing that PolBE is size-extensive.}
\Remove{There is room for investigation of much larger clusters in order to reach a saturation in the binding energy per monomer and other associated errors.}

\subsection{Hydrated ion clusters}

\begin{figure}
     \centering
     \begin{subfigure}[b]{0.49\textwidth}
         \centering
         \includegraphics[width=\textwidth]{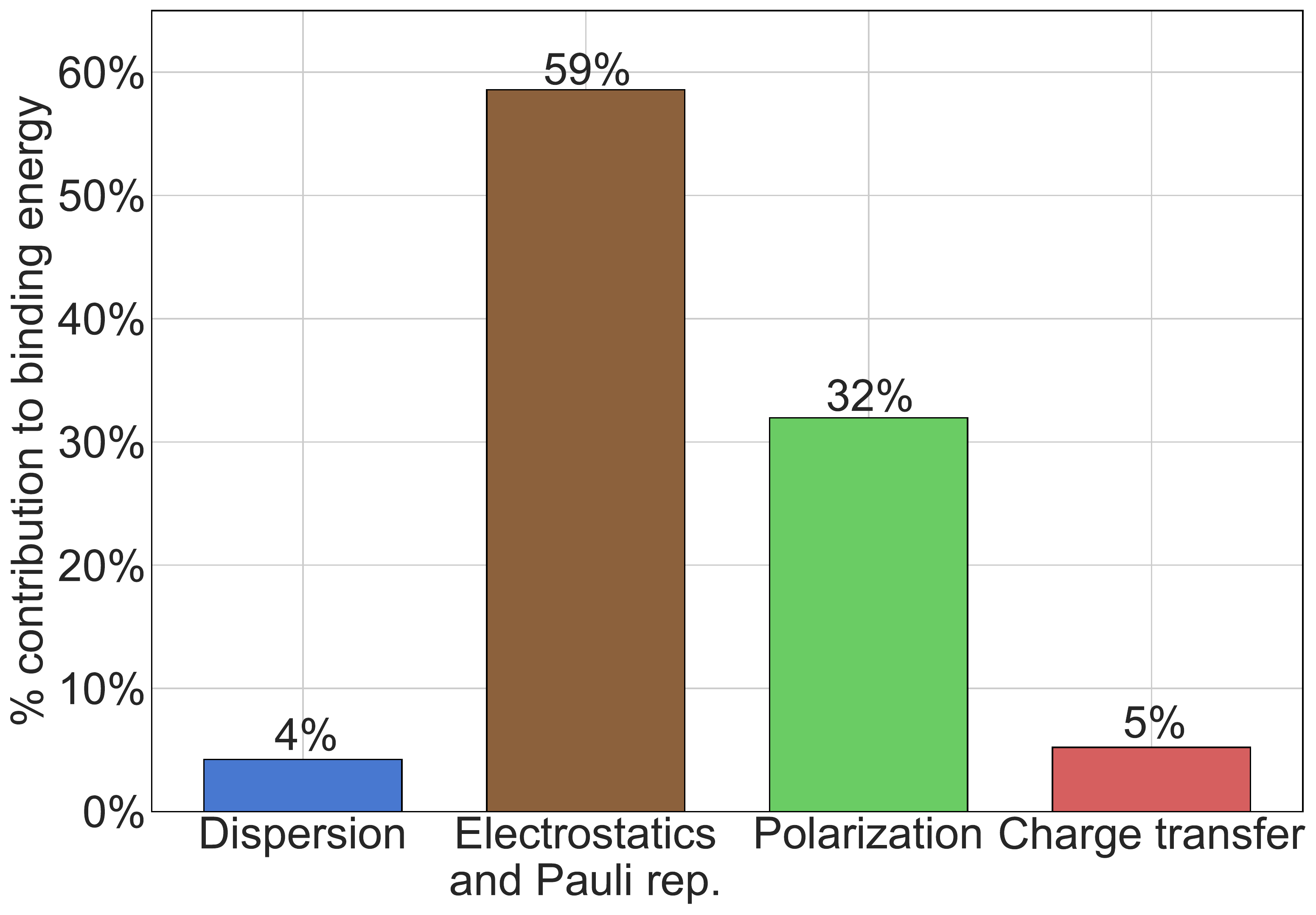}
         \caption{}
         \label{fig:hydratedCa_EDA}
     \end{subfigure}
     \begin{subfigure}[b]{0.49\textwidth}
         \centering
         \includegraphics[width=\textwidth]{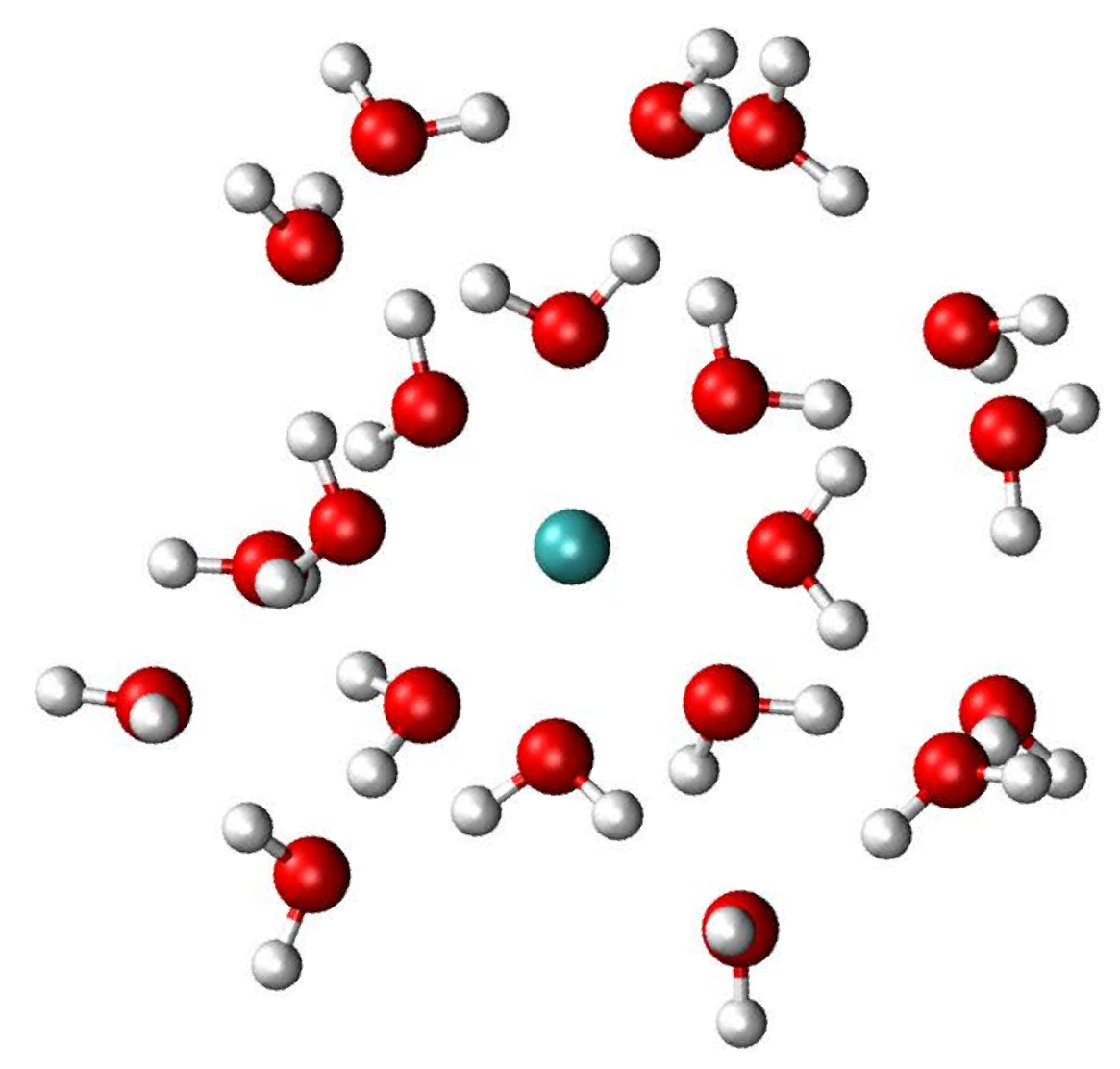}
         \caption{}
         \label{fig:hydratedCa_21mer}
     \end{subfigure}
     \begin{subfigure}[b]{0.49\textwidth}
         \centering
         \includegraphics[width=\textwidth]{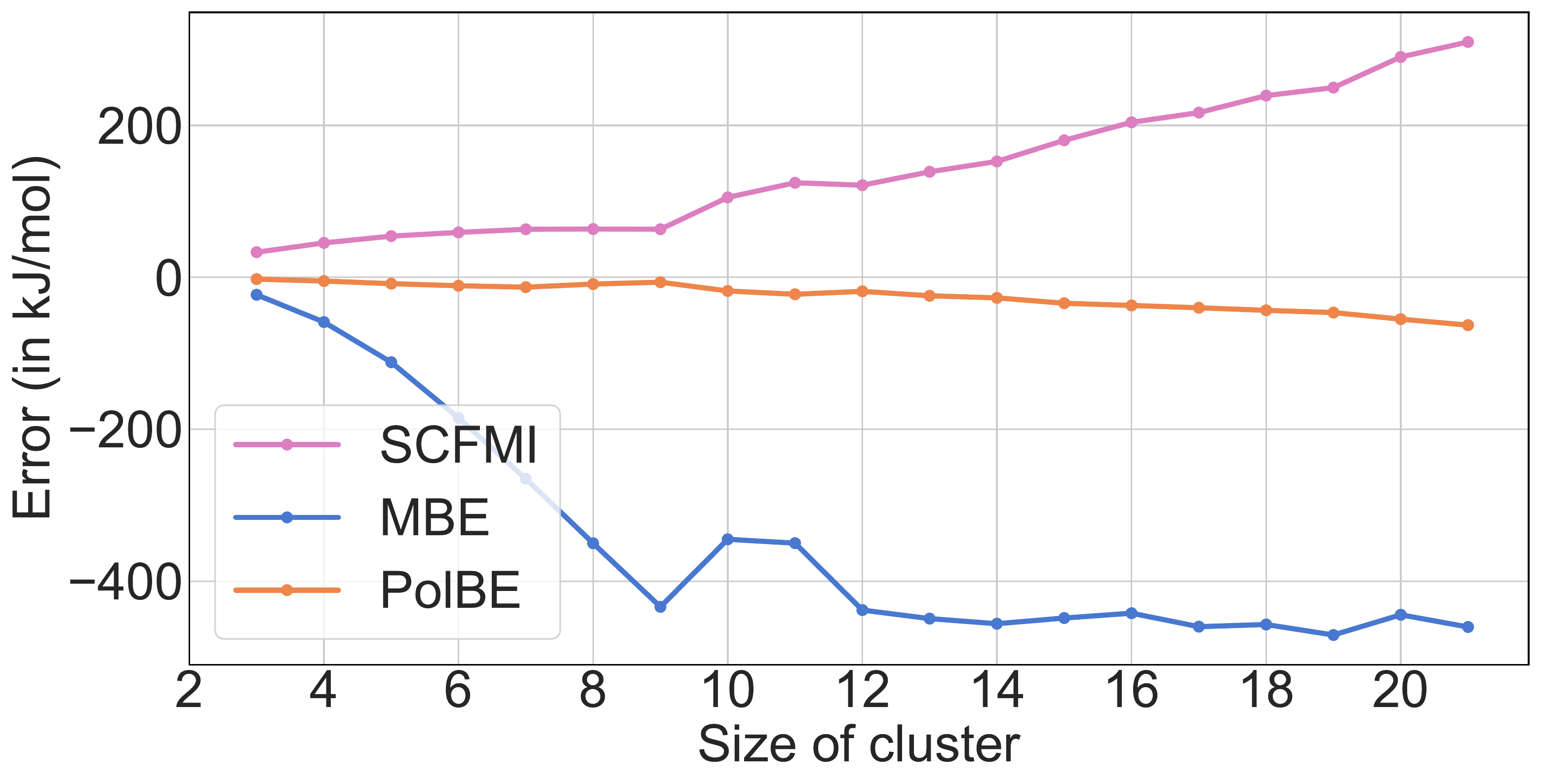}
         \caption{}
         \label{fig:hydratedCa_total}
     \end{subfigure}
     \begin{subfigure}[b]{0.49\textwidth}
         \centering
         \includegraphics[width=\textwidth]{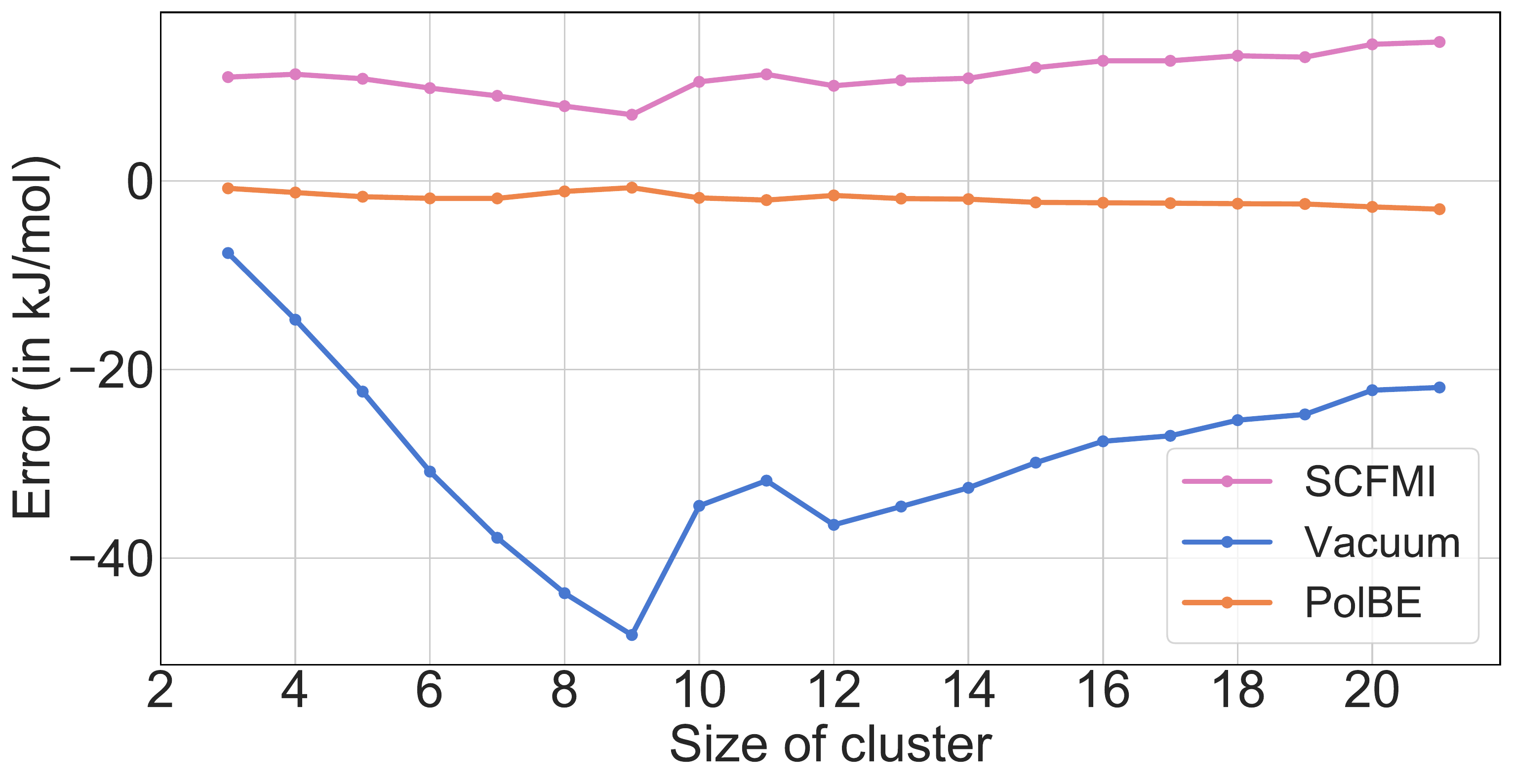}
         \caption{}
         \label{fig:hydratedCa_perfrgm}
     \end{subfigure}
        \caption{(a) Energy decomposition analysis for hydrated \ce{Ca^2+} water clusters (b) Hydrated \ce{Ca^2+} containing 20 water molecules (c) SCFMI, vacuum MBE, and PolBE errors as function of cluster size (d) SCFMI, vacuum MBE, and PolBE errors per fragment as function of cluster size. For comparison, the binding energy of \ce{Ca^2+(H2O)20} is 1935.6 kJ/mol.}
        \label{fig:hydratedCa_graphs}
\end{figure}

\begin{figure}
     \centering
     \begin{subfigure}[b]{0.49\textwidth}
         \centering
         \includegraphics[width=\textwidth]{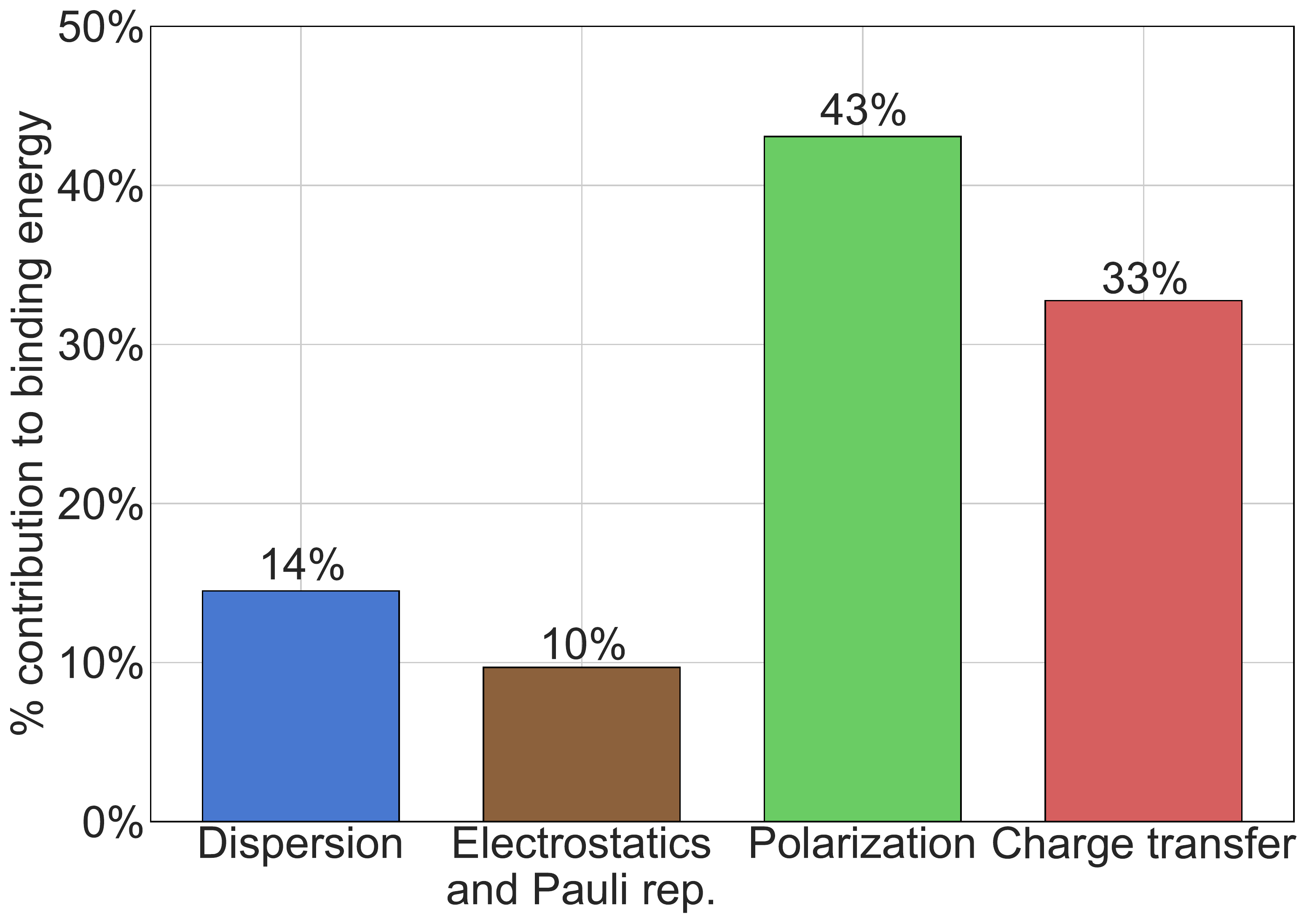}
         \caption{}
         \label{fig:hydratedOH_EDA}
     \end{subfigure}
     \begin{subfigure}[b]{0.49\textwidth}
         \centering
         \includegraphics[width=\textwidth]{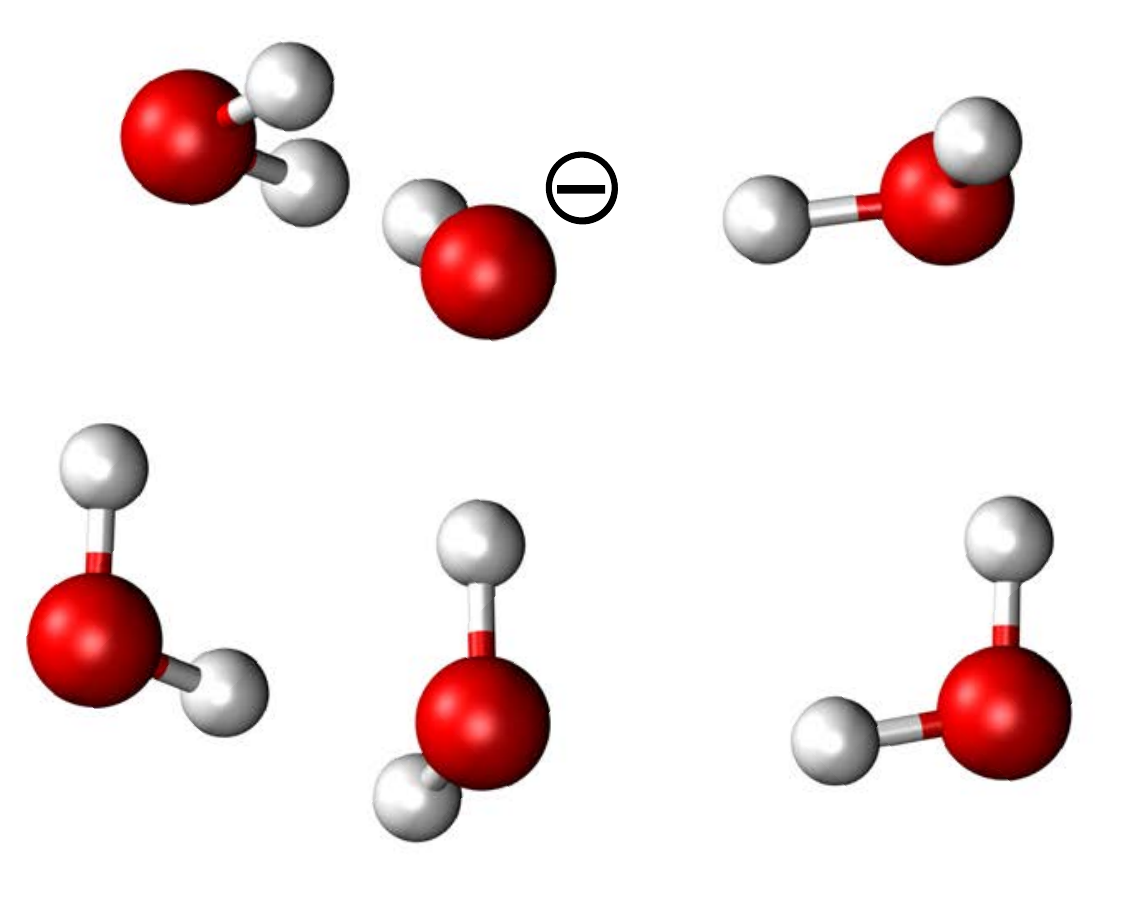}
         \caption{}
         \label{fig:hydratedOH_hexamer}
     \end{subfigure}
     \begin{subfigure}[b]{0.49\textwidth}
         \centering
         \includegraphics[width=\textwidth]{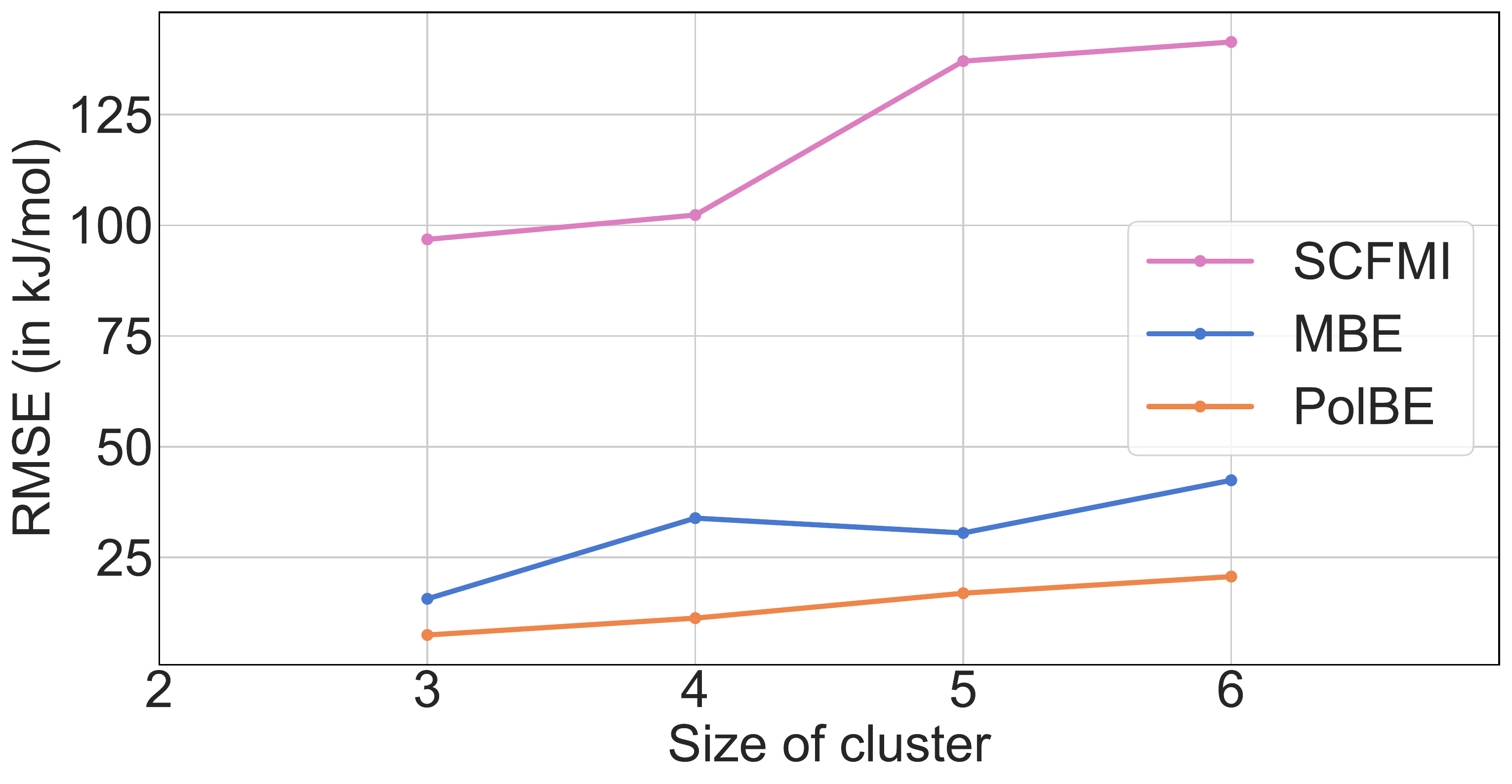}
         \caption{}
         \label{fig:hydratedOH_total}
     \end{subfigure}
     \begin{subfigure}[b]{0.49\textwidth}
         \centering
         \includegraphics[width=\textwidth]{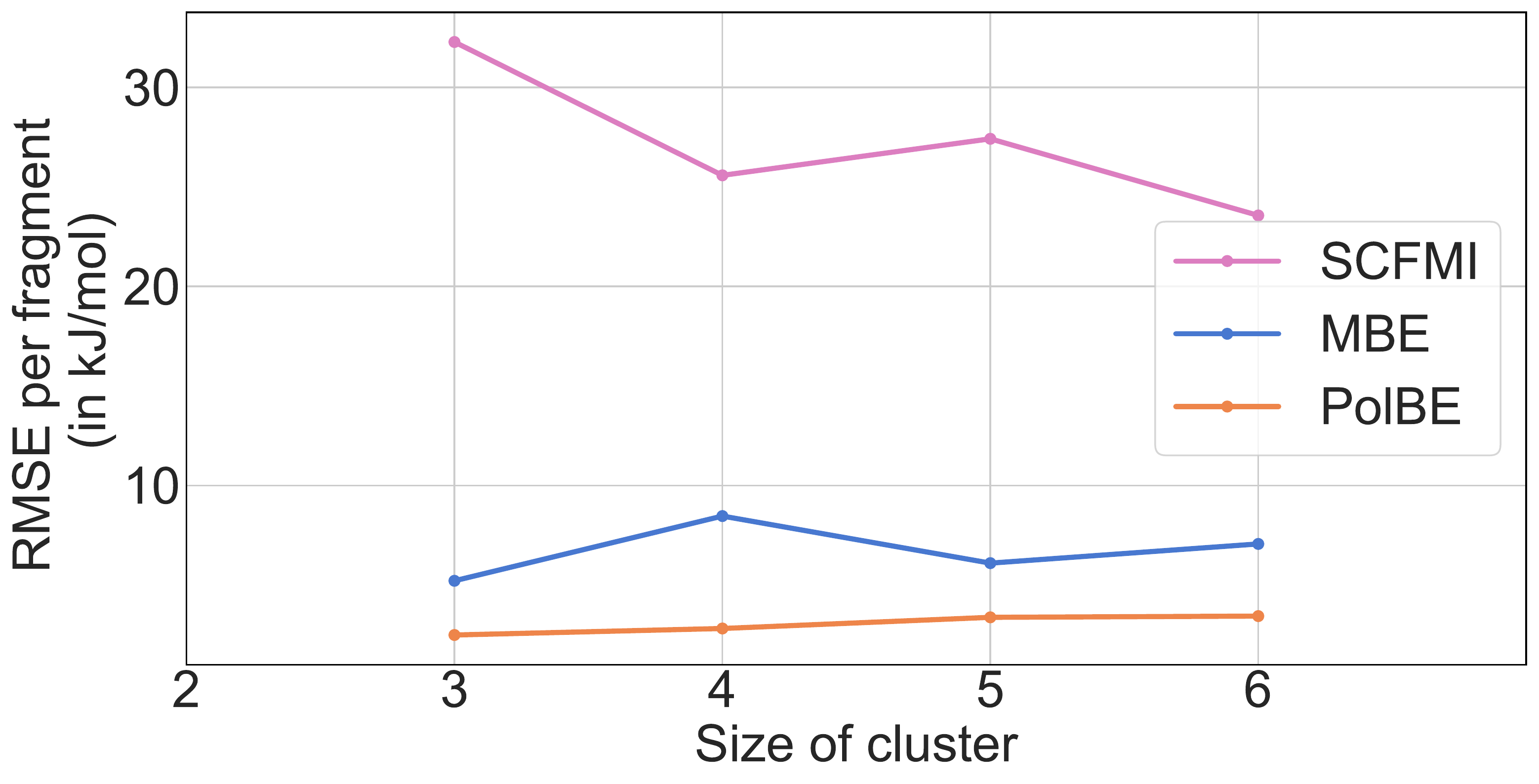}
         \caption{}
         \label{fig:hydratedOH_perfrgm}
     \end{subfigure}
        \caption{(a) Energy decomposition analysis for hydrated \ce{OH-} water clusters (b) Hydrated \ce{OH-} containing 5 water molecules (c) SCFMI, vacuum MBE, and PolBE errors as function of cluster size (d) SCFMI, vacuum MBE, and PolBE errors per fragment as function of cluster size. As multiple conformers of each cluster size were investigated, the error presented here are the root mean square errors of all conformers of a given size. For comparison, the reference RMS binding energy for \ce{OH^-(H2O)5} is 435.6 kJ/mol.}
        \label{fig:hydratedOH_graphs}
\end{figure}

Aqueous ions play a crucial role in many biological processes, and help sustain life on \replacewith{Earth}{earth}.
\eng{While studying bulk aqueous ions has its own set of challenges, studying ion(s) in water clusters serve as a surrogate to understanding their properties.}
The small size of these clusters also enable the usage of accurate and expensive ab initio computational tools and direct comparison with experiment. 
These clusters also serve as models for studying hydration of ions and their coordination shell structure. 
In this section, we present the performance of PolBE and vacuum MBE for cationic water clusters (\ce{Ca^2+}\ce{(H2O)_{n}}, $n=2 - 20$)  and anionic water clusters (\ce{OH-}\ce{(H2O)_{n}}, $n=2 - 6$).
\ce{Ca^2+}\ce{(H2O)_{n}} cluster geometries were taken from Ref.~\citenum{Gonzalez2005} and \ce{OH-}\ce{(H2O)_{n}} geometries are taken from Ref.~\citenum{Egan2018}.

The interaction of ions with water is very strong as can be seen from the large binding energies of the hydrated ion clusters.
The binding energy of \ce{Ca^2+(H2O)20} per fragment is about 92 kJ/mol and that of \ce{OH^-(H2O)5} is 72 kJ/mol.
Decomposition the interaction energy between the \ce{Ca^2+} ion and its surrounding water molecules shows that the interaction is mainly dominated by \replacewith{polarization and the dispersion-free frozen term}{electrostatics and polarization}. 
\remove{The dispersion-free frozen term is large due to permanent electrostatics.}
The bare ion carries a large charge density giving rise to large frozen and polarization terms.
At a frozen density level, this monopole interacts with the surrounding water dipoles which makes permanent electrostatics strong.
These effects are also more long-ranged than in pure water due to the slower decay of monopole-dipole interaction.
The length scale of these effects make\insertnew{s} this \insertnew{class of} system\insertnew{s} \replacewith{not tractable to}{poorly suited for} fragmentation methods, \replacewith{making this system}{and hence} an extreme challenge for PolBE.
The large charge density also creates a strong field polarizing the surrounding water molecules very significantly, thus explaining the large polarization term.

\remove{The biggest notable difference between the binding energies of hydrated ion clusters and regular water clusters is their large interaction binding energies.
\ce{Ca^2+}\ce{(H2O)2} has a binding energy of 433 kJ/mol and \ce{OH-}\ce{(H2O)2} has a binding energy of 231 kJ/mol while a water trimer has an interaction energy of just 63.6 kJ/mol. 
These strong interactions present a formidable challenge to fragmentation-based methods like vacuum MBE. 
Vacuum MBE systematically overbinds clusters of all sizes.
The opposite trend for binding energy per fragment is seen in the case of hydrated ions as it decreases with increasing cluster size.
This is rather expected as the strongest interaction arises from the ions and the ratio of ion to total number of fragments decreases with increasing size of the cluster.}

In the case of hydrated \ce{Ca^2+} ions, the SCFMI wavefunction underestimates the interaction energy as it is unable to capture all the effects of charge transfer. 
The magnitude of underestimation increases from 33 kJ/mol for \ce{Ca^2+}\ce{(H2O)2} to 310 kJ/mol for \ce{Ca^2+}\ce{(H2O)20}.
On the other hand, vacuum MBE error increases from $-$22.9 kJ/mol to $-$460.0 kJ/mol for \ce{Ca^2+}\ce{(H2O)20}.
For the \ce{Ca^2+}\ce{(H2O)20} cluster, this error is about 23\% of the total interaction energy.
Vacuum MBE truncated at the two-body level misses certain important higher order effects.
Vacuum MBE also overestimates the amount of charge transfer at the 2-body level when \insertnew{the} ion comprises one of the two fragments.
Overestimating the interaction energies at a 2-body level causes \insertnew{net} overbinding\replacewith{.}{, while} \replacewith{Neglecting}{neglecting} higher order terms in the MBE causes underestimation of interaction energies.
These two effects compete with each other to cause a temporary saturation in the total vacuum MBE error at around 450 kJ/mol in clusters of sizes 14 to 21.
The saturation of total error with increasing cluster size causes the error per fragment to decrease after the first coordination shell of \ce{Ca^2+} has been saturated.
PolBE performs significantly better by making much smaller total errors, starting from 2.3 kJ/mol for \ce{Ca^2+}\ce{(H2O)2} and increasing steadily to 63.0 kJ/mol for \ce{Ca^2+}\ce{(H2O)20}.
PolBE error, even in these strongly interacting systems, ranges from 0.5 \% to around 3 \% of the total interaction energies\replacewith{.}{,}\remove{PolBE errors are} roughly an order of magnitude smaller than vacuum MBE errors. 
The PolBE error per fragment is only 3.0 kJ/mol for \ce{Ca^2+}\ce{(H2O)20} compared to 1.8 kJ/mol for the water 20-mer.

The first important difference between hydrated \ce{Ca^2+} clusters and hydrated \ce{OH-} clusters is the charge density.
The single negative charge of the \ce{OH-} anion has a much smaller charge density causing smaller total interaction energies.
Comparing the clusters of the same size, hydrated hydroxide ion interaction energies are roughly half of the hydrated calcium ion interaction energies due to the reduced strength of the monopole.
The second difference is the contribution of charge transfer component to the interaction energies.
In hydrated calcium clusters, charge transfer contributes only 5\% to the total interaction energy, while in hydrated hydroxide anion clusters it contributes 33\%.
This is expected because \ce{Ca^2+} is not a good Lewis acid, but \ce{OH-} is a \replacewith{great}{strong} \replacewith{lewis}{Lewis} base and donates electrons to water readily.

As the contribution of charge transfer to interaction energy is significant, the SCFMI wavefunction severely underestimates the interaction energy.
Both vacuum MBE and PolBE overestimate the binding energies for hydrated hydroxide anions, but the errors are shown as positive values as error metric used is root mean square error.
Vacuum MBE binding energy error increases from 15.7 kJ/mol to 42.4 kJ/mol for \ce{OH-}(H\textsubscript{2}O)\textsubscript{5}.
PolBE errors are only about half the magnitude, overestimating the binding energy of \ce{OH-}(H\textsubscript{2}O)\textsubscript{5} by 20.7 kJ/mol.
Similar trends are seen in the binding energy errors per fragment. 

\subsection{Error analysis of different environment models}
\begin{figure}
  \includegraphics[width=\linewidth]{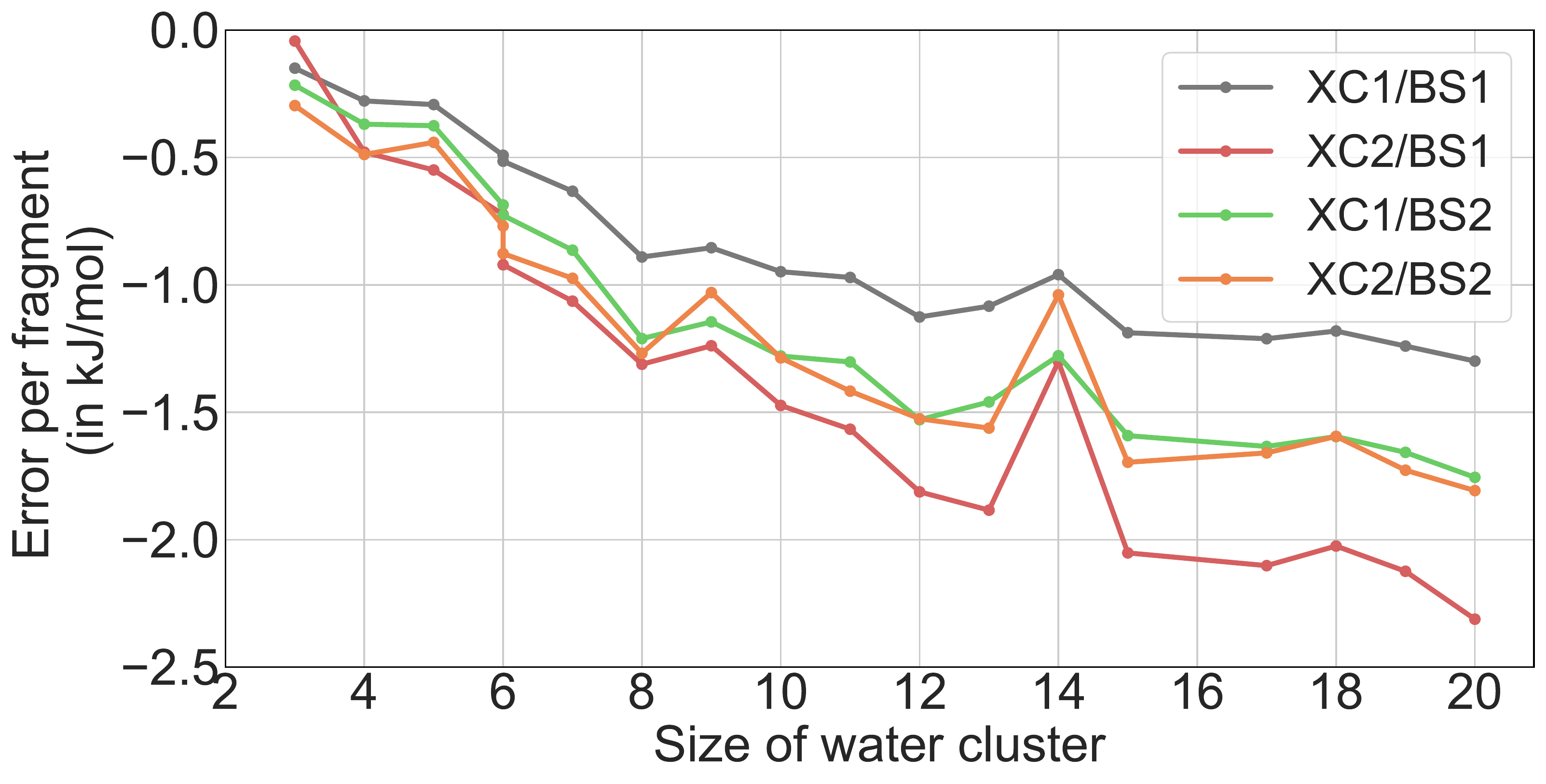}
  \caption{PolBE error per monomer for water clusters as function of size of the cluster with different representations of the environment. In this figure, XC1 is $\omega$B97M-V, XC2 is PBE, BS1 is def2-TZVPPD, and BS2 is def2-SV(P). The system is always treated at XC1/BS1.}
  \label{fig:representation_env}
\end{figure}
The PolBE method uses EMFT to build the Fock matrix and solves the SCFMI equations to obtain ALMOs.
EMFT relies on the fact that an approximate description of the environment (B) is sufficient to accurately capture the essential chemistry occurring in the system (A). 
PolBE relies on this assumption, providing an approximate embedding density for computing 1-body and 2-body terms accurately.
In this section, we discuss the validity of this assumption.
As a mean-field theory, PolBE attempts to represent the environment approximately by using a semi-local functional (XC2) in a much smaller basis set (BS2). 
The PBE functional, while much cheaper to compute than $\omega$B97M-V (XC1), is also grossly inadequate for computing the binding energies of \ce{CO2}, as it does not contain any dispersion and predicts that the \ce{CO2} clusters are unbound.
The def2-SV(P) basis set used for the environment, denoted as BS2, is much smaller than the def2-TZVPPD used for the system.
A single water molecule has 74 basis functions in BS1 and has only 18 basis functions in BS2.
This makes the \replacewith{contribution}{number of contributions} of an environment water molecule to the Fock matrix smaller by a factor of 4 in comparison to the \replacewith{contribution of}{number for} a system water molecule. 

In this section, we analyze the sources of error coming from the functional and basis set approximations used to represent the environment.
In Fig.~\ref{fig:representation_env}, different approximations of the environments are compared in order to isolate the source of error that comes from each of them.
The best representation of the environment uses XC1/BS1 for both the system and the embedding environment.
This treatment of the environment density \eng{on} the same footing as the system does not use EMFT to build the Fock matrix, and is equivalent to SCFMI for each of the 1-body and 2-body terms. 
Therefore, it is the baseline error to compare against (shown as the gray curve in Fig.~\ref{fig:representation_env}).
When the basis set of the environment is changed to BS2 and the functional is fixed at XC1, an additional error of 0.5 kJ/mol per monomer is added at the water 20-mer.
Similarly, changing the environment functional to XC2 while keeping the basis set fixed at BS1 causes the largest error, adding an error of 1 kJ/mol per water molecule to the baseline error.
The method proposed in this paper, which represents the environment using XC2 in a small basis set BS2, adds saturating error of only 0.5 kJ/mol to the baseline error.
This shows that an approximate representation of the environment is sufficient to accurately and efficiently capture the interaction at a 1-body and 2-body level, decomposing the binding energy in an almost pairwise additive fashion.
A systematic analysis of error for different representations of \insertnew{the} environment (different local/semi-local functionals and basis sets) is beyond the scope of this paper.

\subsection{Comparison with existing eMBE methods}
\begin{figure}
  \includegraphics[width=\linewidth]{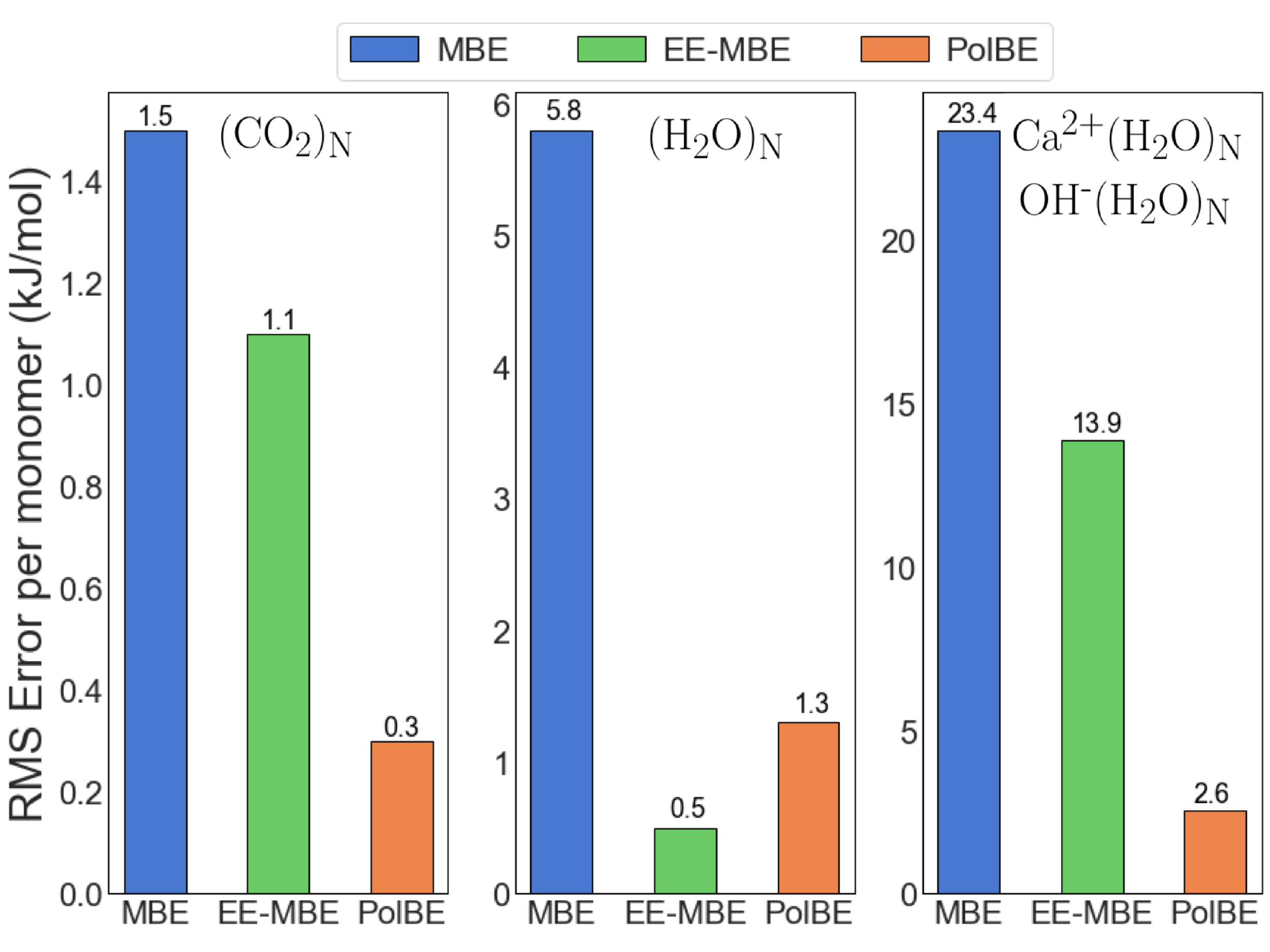}
  \caption{Root Mean Square Error per fragment for all carbon dioxide, water, and hydrated ion clusters. The errors for vacuum MBE, EE-MBE, and PolBE are shown. For \ce{(CO2)_N}}, N ranges from 3 to 16. For \ce{(H2O)_N}, N ranges from 3 to 20. For \ce{Ca^2+(H2O)_N}, N ranges from 2 to 20 and for \ce{OH^-(H2O)_N}, N ranges from 2 to 5.
  \label{fig:EE-MBE}
\end{figure}
Developing different eMBEs has been an active research topic of many research groups for the last couple of decades.
These embedded MBE methods differ only in their description of the environment, ranging from using fixed atom-centered point charges, \cite{Dahlke2007a} using variationally-optimized atom-centered point charges, \cite{Gao2012} to using actual monomer densities.\cite{Fedorov2007}

Of these methods, the Electrostatically Embedded Many-Body Expansion (EE-MBE) \cite{Dahlke2007a} developed by Truhlar and co-workers, is the possibly the simplest and most widely used.
In this method, each k-body term is computed in the presence of atom-centered point charges for the rest of system.
While an approximate representation of the environment \replacewith{is}{can be} sufficient \insertnew{to attain useful accuracy}, EE-MBE oversimplifies it, which causes important interactions like exchange repulsion, coulombic interaction of electron clouds, and polarization to be omitted, consequently leading to poorer performance.
EE-MBE is also not parameter-free as the binding energies depend on the charges as shown by Herbert and co-workers previously. \cite{Richard2014}
Including higher-order terms in EE-MBE can lead to significant precision problems which would require tightening of SCF convergence and integral screening thresholds making them more expensive. \cite{Richard2014a}
Fig. \ref{fig:EE-MBE} shows the errors of EE-MBE performed using CHELPG charges, which are atom-centered charges fitted to reproduce the molecular electrostatic potential at a number of points around the molecule.\cite{Breneman1990}
The CHELPG charges were determined for a water monomer at its equilibrium geometry, and the same set of charges were used for representing all environment water molecules.
In comparison to EE-MBE, PolBE errors are much smaller in the case of \ce{CO2} clusters and hydrated ion clusters.
For all the conformers of hydrated ion clusters studied in this work, the RMS error per fragment of PolBE is only 2.6 kJ/mol whereas that of EE-MBE is 13.9 kJ/mol as seen in Fig.~\ref{fig:EE-MBE}.
For the water clusters investigated, EE-MBE errors are smaller than PolBE errors for the water clusters investigated.
However, the performance of EE-MBE is heavily dependent on the embedding charges used as shown in Table~S8.
For instance, EE-MBE performed using Mulliken charges increases the error per fragment for water molecules to 2.8 kJ/mol.
The errors of EE-MBE also depend on the functional and basis set used to run these calculations.
The performance of PolBE is robust across the range of systems tested in this work and is also parameter-free.
PolBE consistently matches the SCF binding energies computed using XC1/BS1 with modest errors irrespective of the choice of XC1 and BS1.
Different choices for representation of the environment, viz. different XC2 and BS2, does not affect the predicted binding energies \insertnew{too} significantly.

\subsection{Computational timing analysis}

\begin{table}[]
\caption{A table showing the computational cost scaling with system size for construction of Fock matrix and diagonalization steps in SCF and PolBE}
\label{tab:scaling}
\begin{tabular}{|c|c|c|c|c|}
\cline{2-5}
\multicolumn{1}{c|}{}    & \multicolumn{2}{c|}{Fock build}                       & \multicolumn{2}{c|}{Diagonalization}                  \\ \cline{2-5} 
\multicolumn{1}{c|}{}    & \multicolumn{1}{c|}{SCF} & \multicolumn{1}{c|}{PolBE} & \multicolumn{1}{c|}{SCF} & \multicolumn{1}{c|}{PolBE} \\ \cline{1-5} 
2-body screening    &  $\mathcal{O}(F^2)$ & $\mathcal{O}(F^2)+\mathcal{O}(F^3)+\mathcal{O}(F^3)$   &  $\mathcal{O}(F^3)$ & $\mathcal{O}(F)+\mathcal{O}(F)+\mathcal{O}(F)$  \\
No 2-body screening & $\mathcal{O}(F^2)$ & $\mathcal{O}(F^2)+\mathcal{O}(F^3)+\mathcal{O}(F^4)$ & $\mathcal{O}(F^3)$ &   $\mathcal{O}(F)+\mathcal{O}(F)+\mathcal{O}(F^2)$     \\ \hline                 
\end{tabular}
\end{table}

\begin{figure}
     \centering
     \begin{subfigure}[b]{0.49\textwidth}
         \centering
         \includegraphics[width=\textwidth]{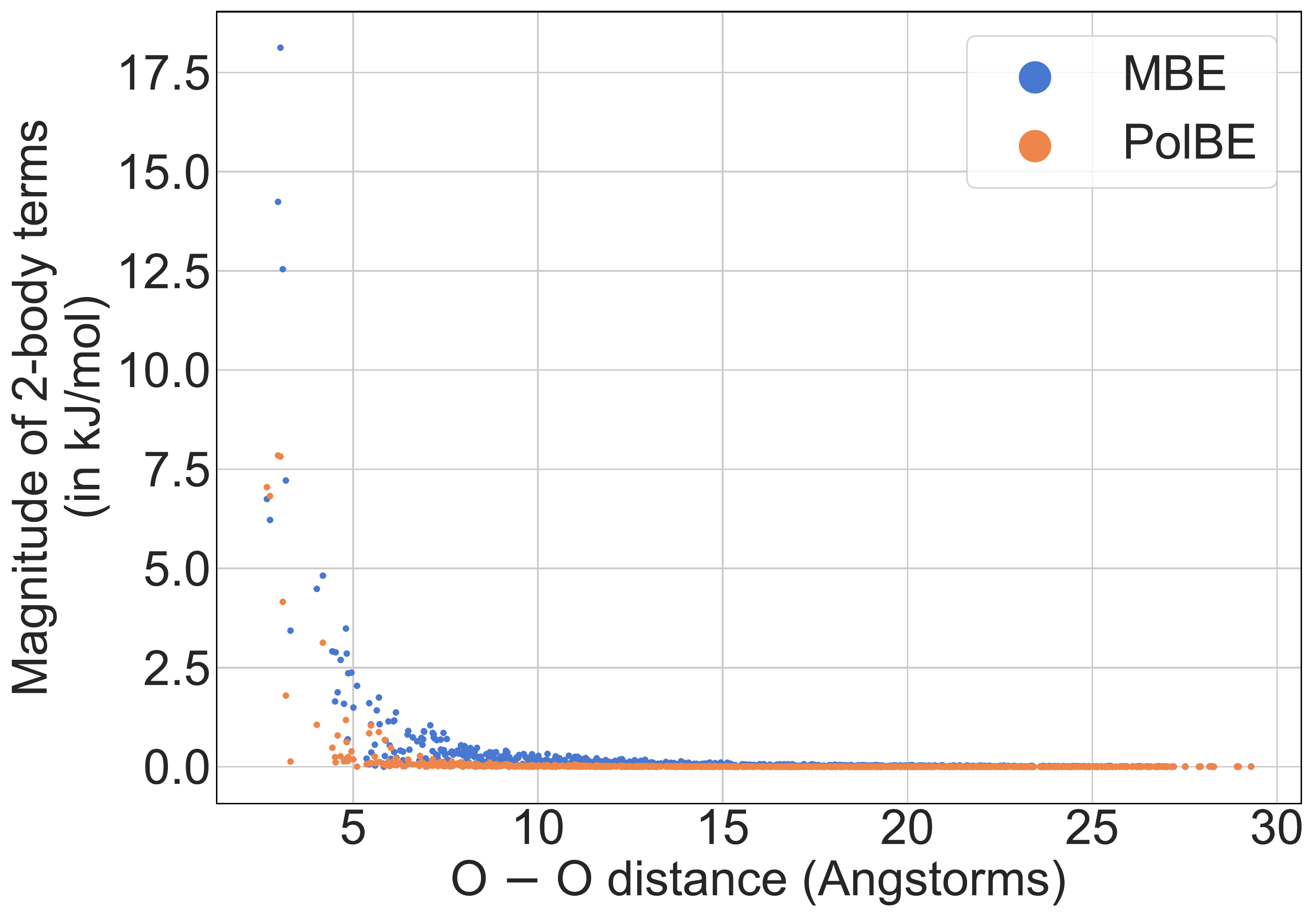}
         \caption{}
         \label{fig:2body_vs_dist}
     \end{subfigure}
     \begin{subfigure}[b]{0.49\textwidth}
         \centering
         \includegraphics[width=\textwidth]{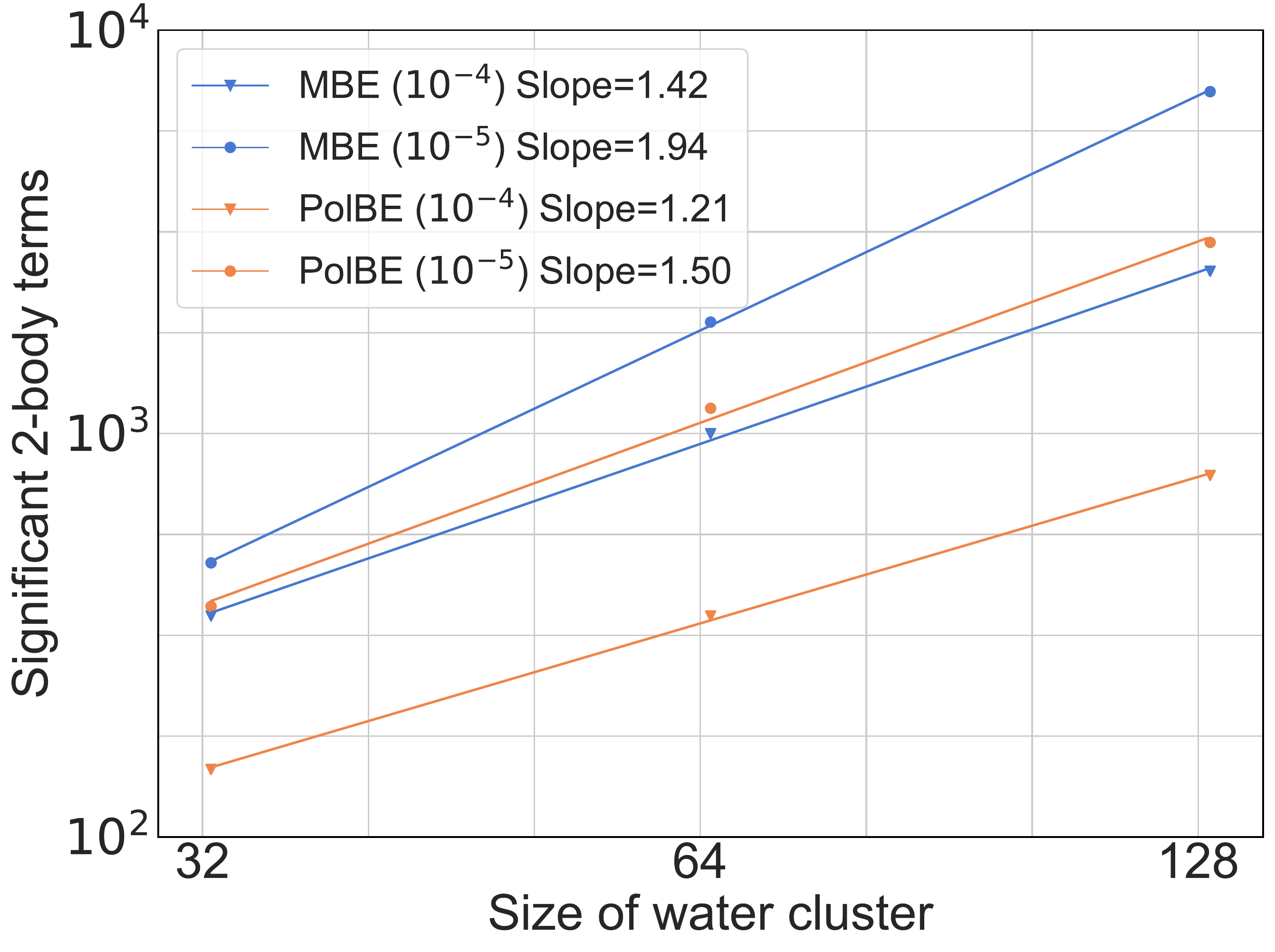}
         \caption{}
         \label{fig:2body_vs_size}
     \end{subfigure}
        \caption{(a) A plot of the magnitude of 2-body terms in MBE and PolBE as a function of the distance between the fragments in a water cluster containing 512 monomers (b) A log-log plot of the number of significant 2-body terms for MBE and PolBE as a function of size of the water cluster. The number of significant terms are indicated determined as the number of terms exceeding the threshold as indicated in parenthesis
        }
        \label{fig:2body}
\end{figure}

\begin{figure}
     \centering
     \begin{subfigure}[b]{0.49\textwidth}
         \centering
         \includegraphics[width=\textwidth]{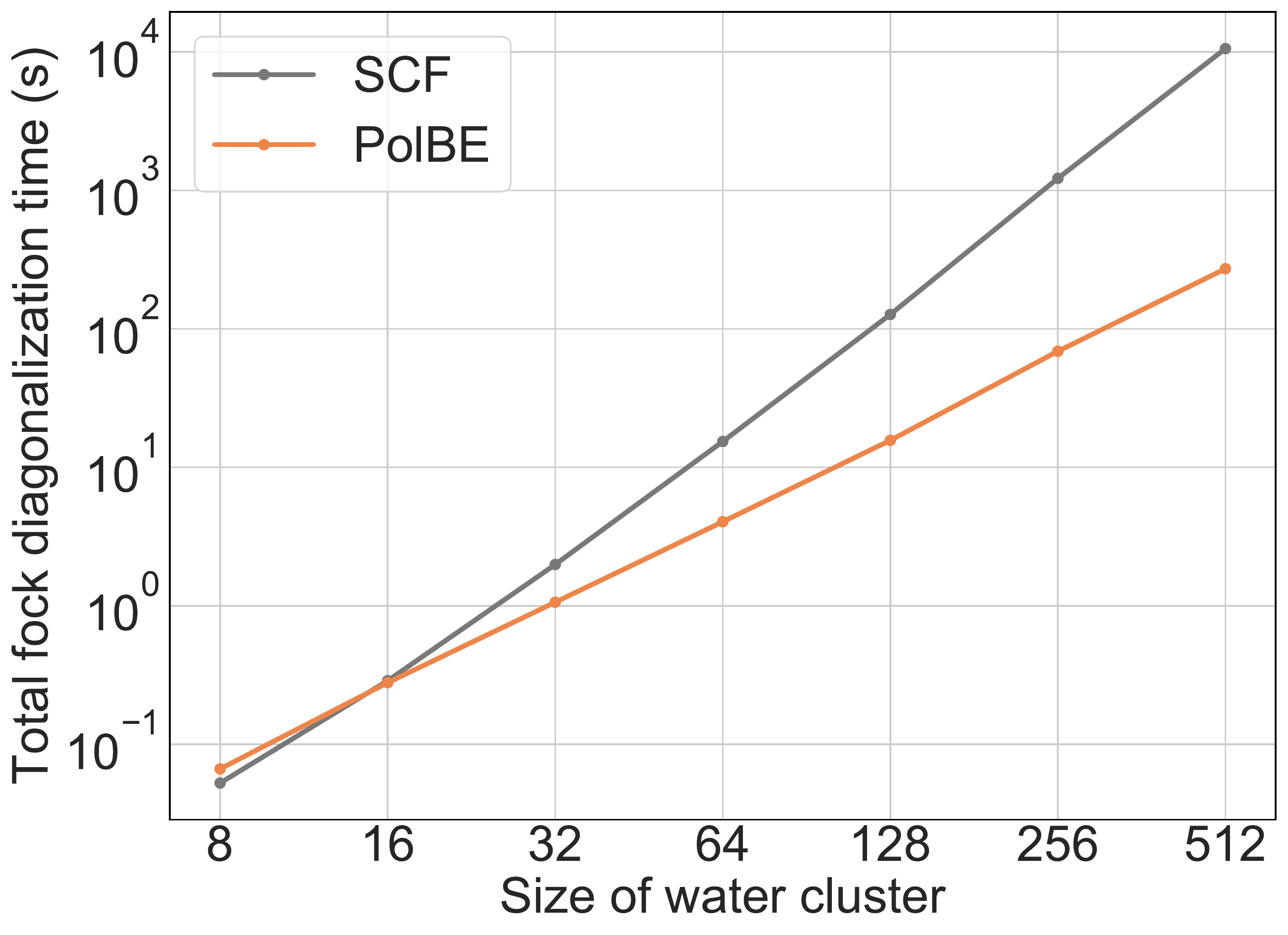}
         \caption{}
         \label{fig:f2m_timing}
     \end{subfigure}
     \begin{subfigure}[b]{0.49\textwidth}
         \centering
         \includegraphics[width=\textwidth]{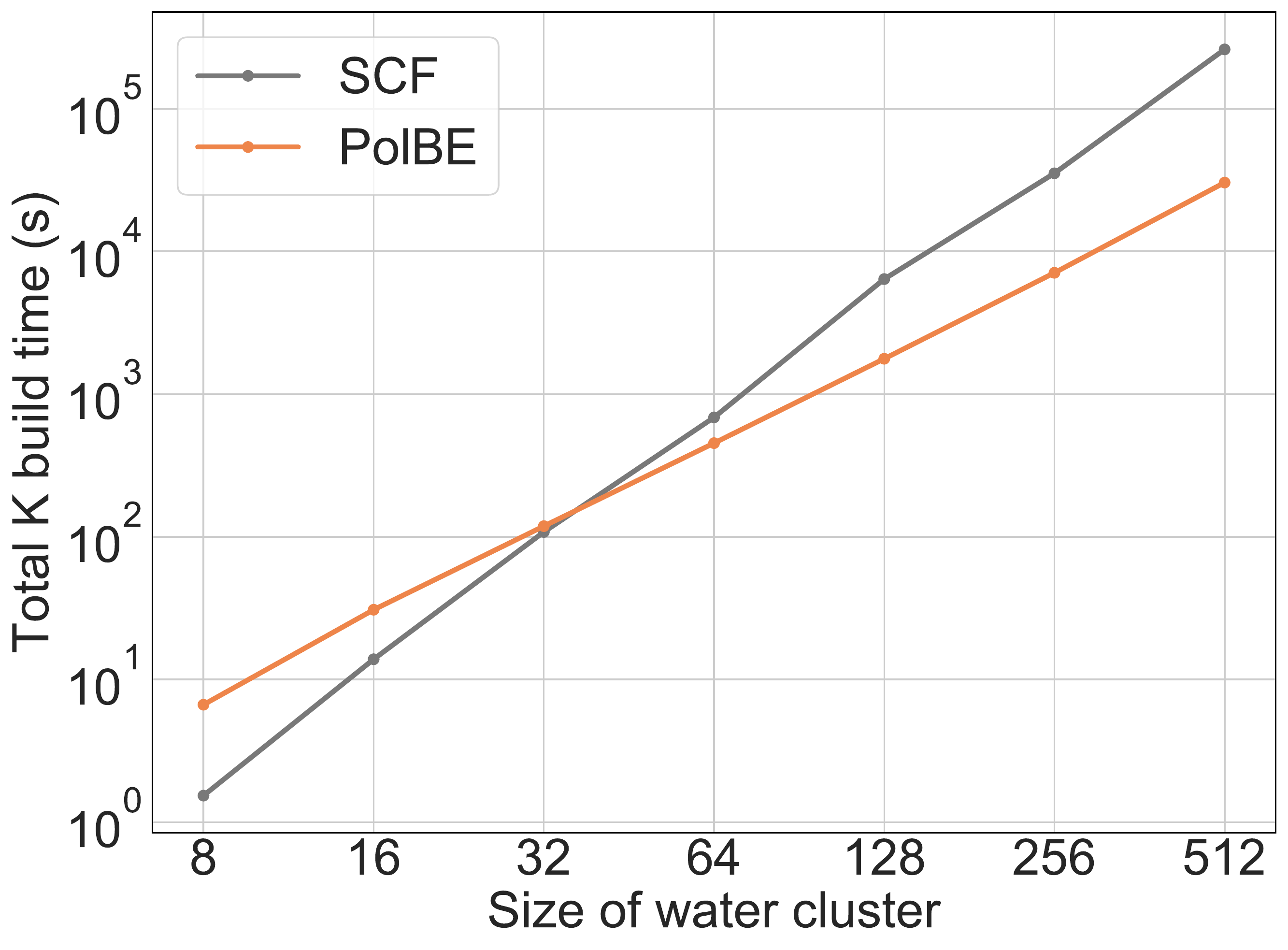}
         \caption{}
         \label{fig:Kbuild_timing}
     \end{subfigure}
        \caption{(a) \insertnew{Wall clock} \replacewith{Time}{time} taken for Fock matrix diagonalization(s) in SCF and PolBE as a function of cluster size (b)\insertnew{Wall clock} \replacewith{Time}{time} taken for construction of exact exchange matrix(ces) for SCF and PolBE as a function of cluster size. $F$ is the number of monomers in the system which indicates system size. \insertnew{Timings reported are the wall clock timings for one iteration of SCF procedure and the wall clock total time for one iteration for all terms in PolBE.}}
        \label{fig:timing}
\end{figure}

\begin{figure}
     \centering
     \begin{subfigure}[b]{0.49\textwidth}
         \centering
         \includegraphics[width=\textwidth]{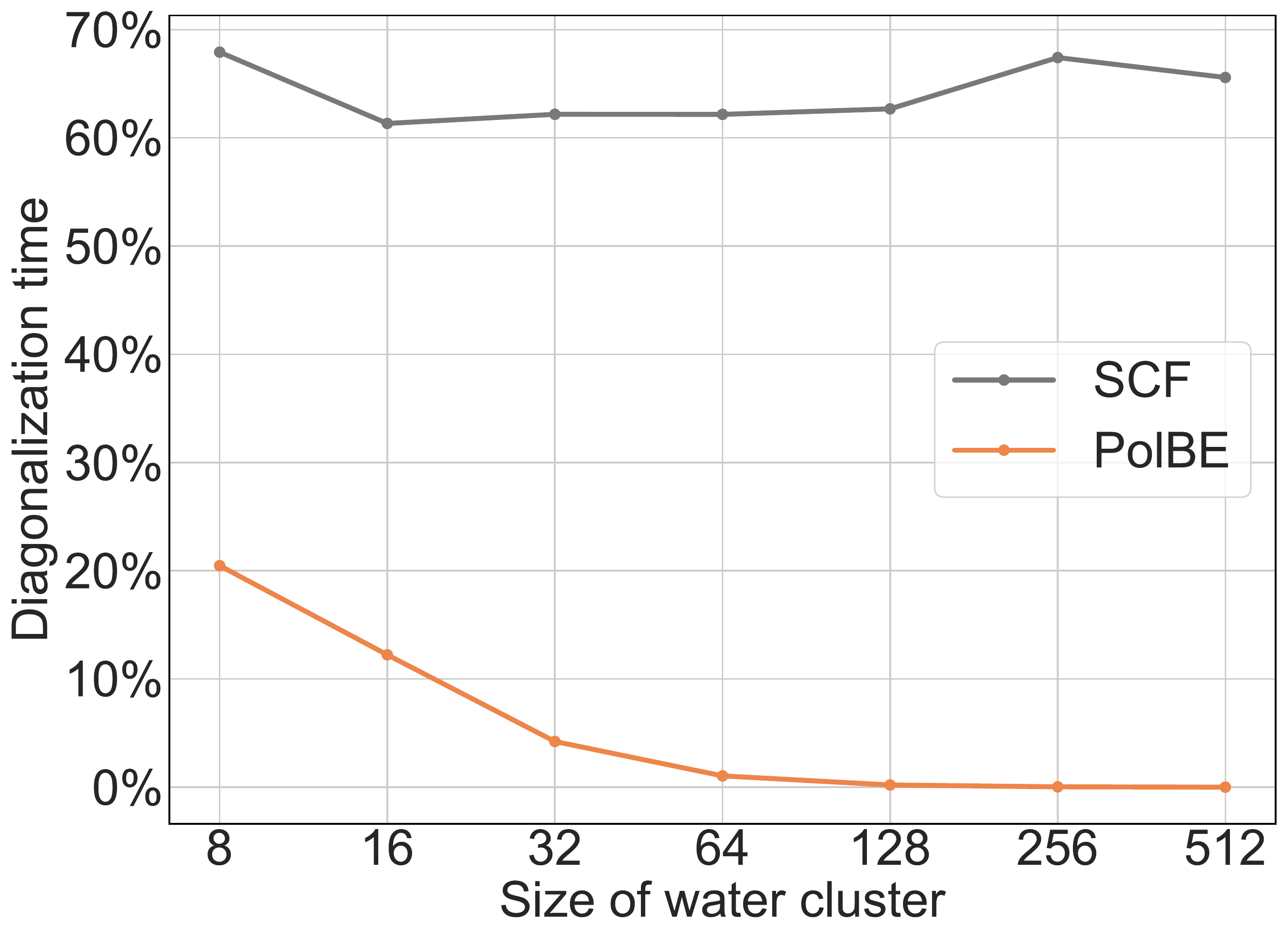}
         \caption{}
         \label{fig:f2m_percent_timing}
     \end{subfigure}
     \begin{subfigure}[b]{0.49\textwidth}
         \centering
         \includegraphics[width=\textwidth]{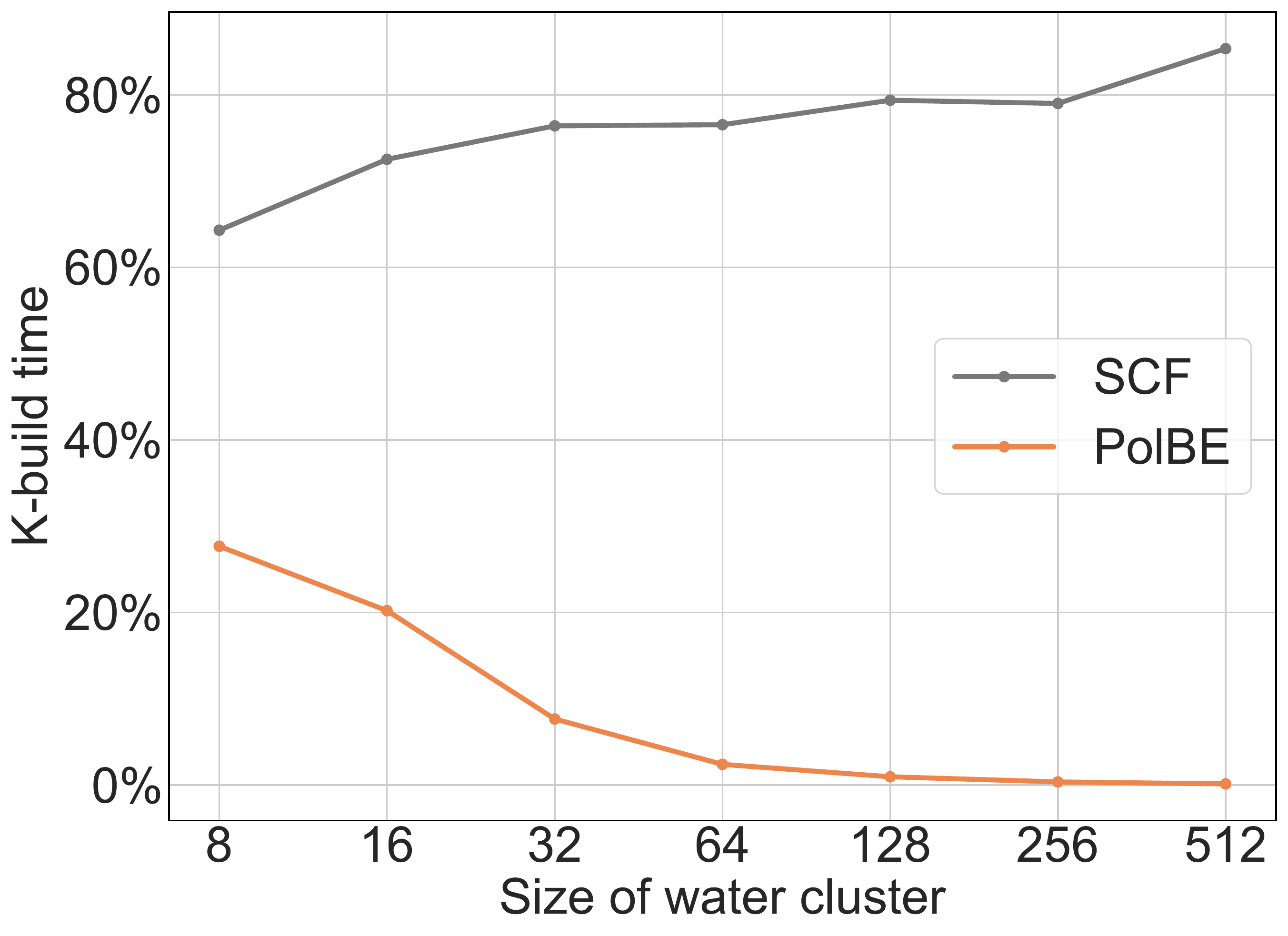}
         \caption{}
         \label{fig:Kbuild_percent_timing}
     \end{subfigure}
        \caption{(a) \insertnew{Wall clock} \replacewith{Time}{time} taken for diagonalization of Fock matrix(ces) as a percentage of total time taken for all linear algebra manipulations for SCF and PolBE as a function of cluster size (b)  \insertnew{Wall clock} \replacewith{Time}{time} taken for construction of the exact exchange bit as a percentage of total time taken for construction of the Fock matrix(ces) for SCF and PolBE as a function of cluster size. \insertnew{Timings reported are the wall clock timings for one iteration of SCF procedure and the wall clock total time for one iteration for all terms in PolBE.}}
        \label{fig:percent_timing}
\end{figure}
The computational expense in the SCF and PolBE procedures can be classified into two categories: diagonalization and Fock build.
In this section, we discuss the computational cost of SCF and PolBE in terms of these two categories.
All timings reported in this section were run on a \insertnew{2.3 GHz 16-core AMD Opteron\texttrademark~processor}.
Timings reported are the wall clock timings for one iteration of SCF\remove{procedure} and the wall clock total time for one iteration for all terms in PolBE.
B3LYP\cite{Becke1993,Lee1988} and PBE\cite{Perdew1996} were used for XC1 and XC2, and def2-SVPD \cite{Rappoport2010a} and STO-3G\cite{Hehre1969} were used for BS1 and BS2 in PolBE.
For the SCF calculation, the reported timings are for running B3LYP/def2-SVPD.
\insertnew{Different density functional and smaller basis sets were used for this part of the timings analysis in order to access larger system sizes.}
Spherical water clusters used in this study were obtained from Ref.\citenum{ErgoSCF}.

The number of terms at each order increases combinatorially with increasing order of the MBE.
While the number of terms in theory increases combinatorially in PolBE as well, not all of them are numerically significant.
This effect can be attributed to embedding, which incorporates physical effects like electrostatics and many-body polarization in the 0-body term as captured at XC2/BS2.
These effects are further corrected using a the high-level functional XC1 in a large basis set BS1 at a 1-body level.
After incorporating 1-body effects, the only physical interactions that are not yet captured are the corrections to the many-body polarization at XC1/BS1 and the effect of charge transfer.
The 2-body terms capture only the effect of 2-body charge transfer and 2-body correction to the polarization, both of which decay rapidly with increasing distance between the pairs.
Fig.~\ref{fig:2body}(a) shows rapid decrease in the magnitude of 2-body terms with distance between the pair.
It also shows that the magnitude of 2-body terms in PolBE are much smaller and decay much faster than 2-body vacuum MBE terms.
This results in the linear increase in the number of 2-body terms in contrast to the quadratic number of all 2-body terms.
Employing a conservative distance-based cutoff will limit the number of 2-body computations to scale linearly with system size.
Considering cluster sizes of 32, 64, and 128 water molecules with a\remove{tight} threshold of $10^{-5}$ Hartrees\remove{(same as the threshold used for converging all energy calculations)}, Fig.~\ref{fig:2body}(b) shows that the number of significant 2-body terms scales sub-quadratically with system size.
Using looser thresholds and/or larger system sizes will show a scaling that is much closer to unity.

Consider a system with $F$ fragments.
Let $N_1$ and $N_2$ represent the total number of basis functions in BS1 and BS2 respectively.
Let $n_1$ and $n_2$ represent the numbers of basis functions per monomer in BS1 and BS2 respectively.
The cost of Fock matrix construction for SCF scales as $\mathcal{O}(N_1^2)$ leading to\remove{a}$\mathcal{O}(F^2)$ scaling with system size.\cite{Almlof1982}
PolBE involves construction of three types of Fock matrices: (a) One Fock matrix at XC2/BS2 for $V_0$ (b) \insertnew{A} \replacewith{Linear}{linear} number of Fock matrices for the $V_1$ term (c) \insertnew{A} \replacewith{Linear}{linear}\remove{number}or quadratic number of Fock matrices for \insertnew{the} $V_2$ term depending \replacewith{of}{on} whether screening is employed or not employed\remove{respectively}.
For each 1-body term, the size of the XC2 Fock matrix is roughly $N_2$.
$V_1$ Fock matrices would also involve the computation of two additional Fock contributions from XC2 and XC1, but their size is limited to $n_1$ and does not scale with system size.
The total cost of computation of the $V_1$ terms scales as $F \times \mathcal{O}(N_2^2)$.
Each Fock matrix for the $V_2$ term also exhibits similar computational costs.
Summing up, the cost of \replacewith{construction}{constructing} all the Fock matrices scales as $\mathcal{O}(F^2)+\mathcal{O}(F^3)+\mathcal{O}(F^3)$ after employing 2-body screening.

The density update step, which involves diagonalization of the effective hamiltonian matrix, scales \replacewith{at the}{as} $O(M^3)$, where M is the size of the system.
For large system sizes, diagonalization of the Fock matrix becomes the most expensive step computationally.
Diagonalization accounts for the majority of the computational cost of all linear algebra manipulations, constituting more than 60\% of the cost across all fragment sizes as shown in Fig.~\ref{fig:percent_timing}(a).
\remove{Different diagonalization-free SCF methods have been proposed; SCFMI is one of them. \cite{Khaliullin2006, Khaliullin2013}
SCFMI involves diagonalization of only the projected Fock matrices which are of fragment dimension (either $n_1$ or $n_2$).}
PolBE, which is based on the SCFMI wavefunction, removes this diagonalization bottleneck for large systems as shown in Fig.~\ref{fig:percent_timing}\insertnew{(a)}.
As the cost of diagonalization scales as the third power of the dimension of the matrix, density update step in SCF scales as $\mathcal{O}(N_1^3)$.
In PolBE at the $V_0$ level, the total cost of diagonalization scales as $F \times \mathcal{O}(n_2^2)$.
As $n_2$ is constant, the cost of diagonalization for $V_0$ term scales as $\mathcal{O}(F)$.
For each term in $V_1$ and $V_2$, only the \replacewith{molecular orbitals}{MOs} of the monomer and dimers are updated while the \replacewith{molecular orbitals}{MOs} of the environment are fixed.
This requires diagonalization of projected Fock matrices \replacewith{of the}{with} dimension\remove{of} $n_1$ and $2n_1$ for the $V_1$ and $V_2$ terms respectively.
In conclusion, the diagonalization step scales as $\mathcal{O}(F)+\mathcal{O}(F)+\mathcal{O}(F)$ with system size.
The total cost of diagonalization in SCF quickly exceeds that of PolBE for cluster sizes of 16 water molecules (624 basis functions) as shown in Fig.~\ref{fig:timing}.

\remove{The construction of }Fock matrix \insertnew{construction},\remove{which} comprises\remove{of computing the}coulomb, XC and exact exchange \replacewith{bits}{contributions}, \insertnew{and} scales as the 2\textsuperscript{nd} power of its size.
\remove{The} \replacewith{exact}{Exact} exchange\remove{bit} dominates the computational cost\remove{of building the Fock matrix} as shown in Fig.~\ref{fig:percent_timing}(b).
Hybrid density functionals, known to alleviate the self-interaction energy problem, provide a more reliable and accurate prediction of system properties, including non-covalent interaction energies. \cite{Mardirossian2017}
However, addition of exact exchange \replacewith{to the}{relative to} pure density functionals makes them more expensive to compute. \cite{Manzer2015,Manzer2015a,Ochsenfeld1998}
Different methods, ranging from screening based on locality to density-fitting have been proposed for reducing the computational cost of forming the Fock matrix.\cite{Ochsenfeld1998}
However, computation of exact exchange has remained the bottleneck step.
\remove{Treating dispersion interaction correctly with the VV10 non-local correlation functional also adds to the cost of computation.} 
As shown in Fig.~\ref{fig:percent_timing}(b), construction of the exact exchange part of the Fock matrix takes up upwards of 60\% of the total Fock construction time.
PolBE, which embeds hybrid DFT in semi-local DFT, requires construction of only the AA block of the exact exchange matrix which is much smaller than the entire Fock matrix.
The size of the AA block is only $n_1$ for each 1-body term and $2n_1$ for each 2-body term.
The total SCF K-build time exceed the total K-build time for all the $^FC_1+^FC_2$ terms in PolBE at 32 water molecules (1248 basis functions) as shown in Fig.~\ref{fig:timing}(b).
PolBE thus \replacewith{removes}{accelerates} the computational bottleneck step in both linear algebra manipulations and Fock construction.
In terms of the total computational cost, PolBE costs more than a full SCF calculation owing to the large number of Fock constructions at a 2-body level.
However, all of each of these computations are trivially parallel as described below.

\begin{figure}
  \includegraphics[width=\linewidth]{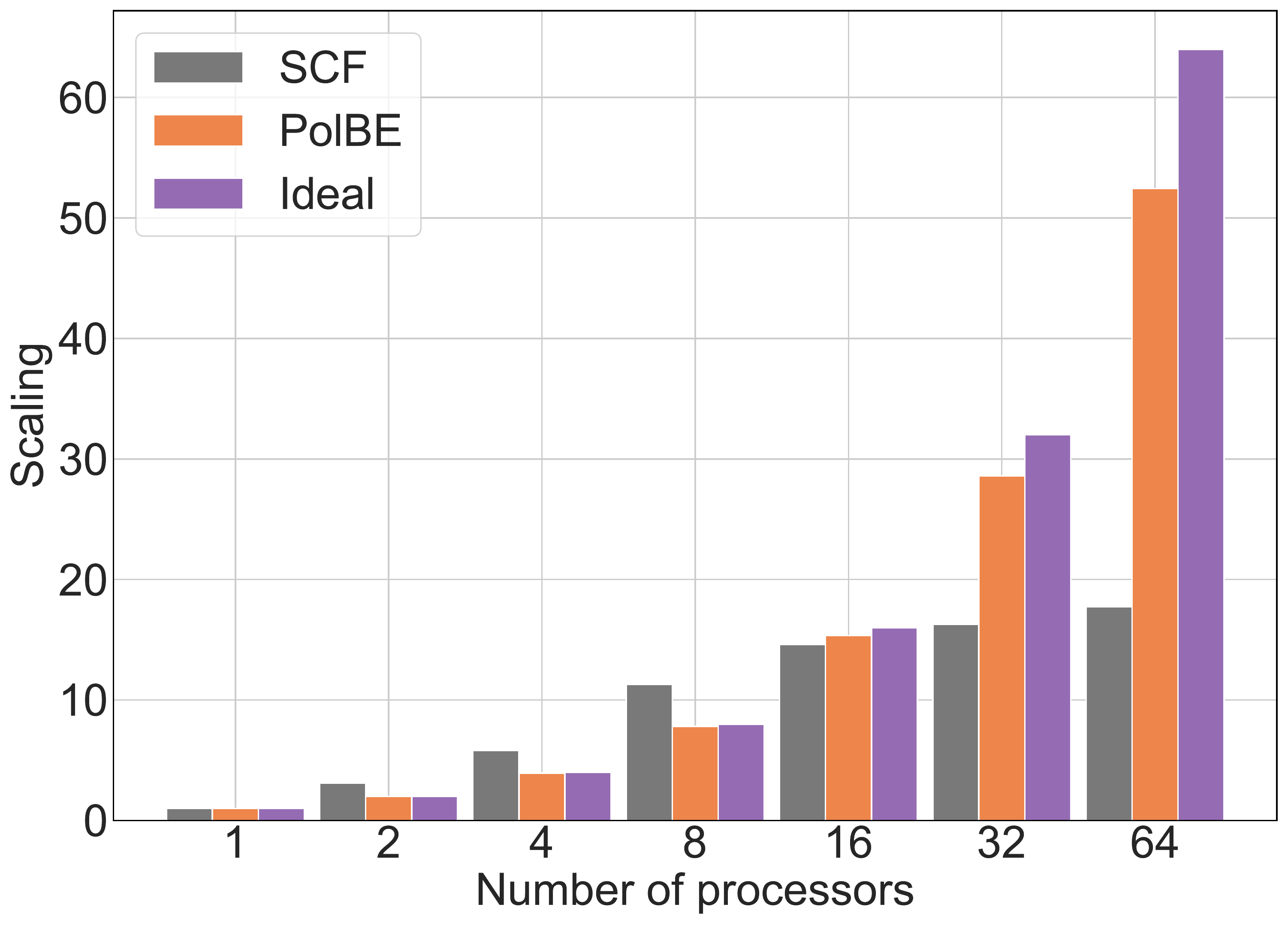}
  \caption{Scaling of SCF and PolBE with increasing number of processors for water 20-mer. Computations were performed using $\omega$B97M-V/def2-TZVPPD in PBE/def2-SV(P) for PolBE and $\omega$B97M-V/def2-TZVPPD for SCF.}
  \label{fig:parallelizability}
\end{figure}

A practical challenge in most DFT software packages is to be able to take full advantage of the multiple\remove{hundreds of} cores available in modern day high performance computing architectures. \cite{Hu2015}
PolBE provides a \insertnew{very simple} way to take advantage of this.
Once the embedding density has been formed, PolBE involves computation of $^FC_1+^FC_2$ terms, all of them\remove{being} completely independent of each other.
This property of PolBE, inherited from the many-body expansion, makes it trivial to distribute these jobs among any number of given processors making PolBE ``embarrassingly parallel.''
As shown in Fig.~\ref{fig:parallelizability},\remove{the}PolBE \replacewith{scales close to that of ideal parallelizability}{for water clusters exhibits nearly ideal parallel scaling}.
In principle, one can achieve even better scaling with PolBE by conceiving smarter master-slave process which are aware of the approximate time it would take to run each of the jobs \textit{a priori}.

\newpage

\section{Conclusions and Outlook}
In this work, we \replacewith{investigate}{investigated} the importance of accurately representing the environment in order accelerate the convergence of the many-body expansion.
In particular, we found that capturing the effects of the environment using EMFT while partitioning the fragments using SCFMI accurately captures the many-body cooperativity effects commonly found in condensed phase systems.
We have shown that the PolBE method proposed in this paper reproduces binding energies of a wide range of molecular clusters of various sizes and consisting of different interaction mechanisms.
PolBE consistently performs better than other popular embedded many-body expansion methods  EE-MBE.
The computational cost of PolBE and its ability to remove the computational bottleneck steps \replacewith{is}{was analyzed and} demonstrated.

\Insertnew{Comparison of the performance of PolBE to other embedded many-body expansion methods\cite{Bygrave2012} would be of significant interest.}
\insertnew{The implementation of PolBE reported here can potentially be improved using various strategies. }\remove{Investigation of} \replacewith{larger}{Larger} systems can be made computationally more tractable by using a real-space cutoff for computing two-body terms. 
These cut-offs \replacewith{should}{must} depend on the dominant long-range interactions present in the system and the accuracy desired. 
Such cut-offs have not been investigated in this paper and are an interesting topic for future study.
As PolBE is an MBE based method, its accuracy can be systematically improved by including higher order terms in the expansion.
While the number of higher order terms increases with each order, some screening of important terms, either based on real-space distances or on the magnitude of lower terms, can be employed.
The description of the environment can be made\remove{further} coarser \insertnew{to achieve additional speedups}. 
PolBE, as suggested in this work, uses different basis sets and functionals for the system and environment.
The extension of PolBE to use different DFT quadrature grids for the system and the environment is straightforward.
Similar ideas can be extended to use different density fitting basis sets for the coulombic interaction term.
\Insertnew{The ALMO-based  representation of the system used in PolBE yields fragment-specific occupied and virtual orbitals.
These fragment-specific virtuals, which are both fragment-sparse and smaller in number in comparison to the virtual space of the whole system, provides a compact representation of the fragment virtual space which can be exploited in the extension of PolBE to double hybrids density functionals\cite{Mardirossian2018} and to  correlated wavefunction methods like MP2, orbital-optimized MP2,\cite{Lochan2007, Stuck2013, Lee2018, Lee2019, Lee2019a, Bertels2019} and coupled cluster theories.\cite{Raghavachari1989} }
Extension of these ideas to correlated wavefunction methods is currently being investigated.

\newpage

\section{Supplementary Material}
See supplementary material for EDA analysis, EMFT EX0 numerical instability, numerical values of binding energy, binding energy errors, charges used for EE-MBE, and details of the computational timings analysis.

\section{Acknowledgements}
This work was supported by the National Science Foundation (NSF) under award CHE-1665315. 
The authors declare the following competing financial interest(s): M. H.-G. is a part owner of Q-Chem, Inc.

\newpage

\bibliography{references}
\bibliographystyle{achemso}
\end{document}